\documentstyle[emulateapj,psfig,url]{article}

\makeatletter

\newenvironment{inlinefigure}{%
\def\@captype{figure}%
\noindent\begin{minipage}{0.999\linewidth}\begin{center}}
{\end{center}\end{minipage}\smallskip}
\makeatother

\lefthead{Cowie, Barger, \& Trouille}

\begin{document}
\title{Measuring the Sources of the Intergalactic Ionizing Flux\altaffilmark{1}} 
\author{
L.~L.~Cowie,\altaffilmark{2}
A.~J.~Barger,\altaffilmark{2,3,4}
L.~Trouille\altaffilmark{3}
}

\altaffiltext{1}{Based in part on data obtained at the W. M. Keck
Observatory, which is operated as a scientific partnership among the
the California Institute of Technology, the University of
California, and NASA and was made possible by the generous financial
support of the W. M. Keck Foundation.}
\altaffiltext{2}{Institute for Astronomy, University of Hawaii,
2680 Woodlawn Drive, Honolulu, HI 96822.}
\altaffiltext{3}{Department of Astronomy, University of
Wisconsin-Madison, 475 North Charter Street, Madison, WI 53706.}
\altaffiltext{4}{Department of Physics and Astronomy,
University of Hawaii, 2505 Correa Road, Honolulu, HI 96822.}

\slugcomment{Accepted by The Astrophysical Journal}

\begin{abstract}
We use a wide-field (0.9~deg$^2$) X-ray sample 
with optical and GALEX ultraviolet observations to
measure the contribution of  Active Galactic Nuclei (AGNs)
to the ionizing flux as a function of redshift. Our
analysis shows that the AGN contribution 
to the metagalactic ionizing background peaks at around $z=2$.
The measured values of the ionizing background from
the AGNs are lower than previous estimates 
and confirm that ionization from AGNs 
is insufficient to maintain the observed ionization of the
intergalactic medium (IGM) at $z>3$.
We show that only sources with broad lines in their optical
spectra have detectable ionizing flux and that the 
ionizing flux seen in an AGN is not correlated with its X-ray 
color. We also use the GALEX observations of the GOODS-N region to
place a $2\sigma$ upper limit of 0.008 on the
average ionization fraction
$f_{\nu}(700~{\rm \AA})/f_{\nu}(1500~{\rm \AA})$
for 626 UV selected galaxies in the 
redshift range $z=0.9-1.4$.
We then use this limit to estimate  an upper bound to the galaxy
contribution in the redshift range $z=0-5$.
If the $z\sim1.15$ ionization fraction is 
appropriate for higher redshift galaxies,
then contributions from the galaxy population are also
too low to account for the IGM ionization at the highest
redshifts ($z>4$).
\end{abstract}


\section{Introduction}
\label{secintro}

One of the most important parameters in the cosmological modeling 
of galaxies and the intergalactic gas is the level of the ionizing 
radiation in the universe. This is often referred to as the 
metagalactic ionizing background. The metagalactic ionizing 
background is usually computed using composite quasar spectra convolved 
with the quasar luminosity function (e.g., Haardt \& Madau 1996;
Madau et al.\ 1999; Haardt \& Madau 2001; Meiksin 2005).
Although the relative contributions of the
galaxy population and the Active Galactic 
Nucleus (AGN) population are very poorly determined, based on the
ionization stages seen in the intergalactic gas, we suspect that AGNs 
dominate the production below redshifts of about three.
At higher redshifts recent estimates suggest that the AGNs are unable 
to account for the ionizing flux (e.g., Bolton et al.\ 2005; Meiksin 2005).
In this paper we will confirm this result and show that 
the AGN contribution is even lower than previously calculated.

It is usually assumed that the ionizing flux at these
higher redshifts is predominantly
produced by galaxies. However, there is very little evidence that 
enough ionizing photons escape from the galaxies for this to be 
true. Most measurements for local and $z=1$ galaxies have given 
only upper limits on the escape fractions (Leitherer et al.\ 1995; 
Malkan et al.\ 2003; Siana et al.\ 2007), as have
recent measurements at $z=3$ (Giallongo et al.\ 2002;
Inoue et al.\ 2006; Fernandez-Soto et al.\ 2003).
However, there is some counter evidence for significant escape fractions,
though it is unclear how representative these galaxies are of the general
population. Bergvall et al.\ (2006) claimed a significant escape fraction in 
the local blue compact galaxy Haro~11. 
Shapley et al.\ (2006) detected ionizing fluxes from two $z=3$
Lyman Break Galaxies (LBGs), though these measurements were at a 
much lower level than a previous measurement made by 
Steidel et al.\ (2001). Iwata et al. (2008), using a narrow
band filter technique, found ionizing
radiation in seven $z>3$ LBGs  and 10 $z>3$ Lyman alpha emitters
out of a sample of 198 $z>3$ galaxies in the SSA22 field.
However, their analysis of the level of contamination by foreground
galaxies is quite crude, and it is also possible that at
least some of the detections correspond to objects with misidentified
redshifts. Paradoxically, Iwata et al.\ found a highly significant null detection
for the object SSA22a-D3, which is the brighter of the two
objects  for which Shapley et al.\ (2006) claimed an ionizing flux detection.

Here we use UV images from the Galaxy 
Evolution Explorer (GALEX) mission (Martin et al.\ 2005) to 
measure the actual ionizing fluxes from very large samples 
of both galaxies and AGNs. As we illustrate in Figure~\ref{filter_fig},
the 1528~\AA\ filter on GALEX samples
the ionizing region of the spectrum at $z=1$. The 2371~\AA\ filter
samples it at $z=2$. This avoids issues of the calibration
of the spectra and also of the selection functions present in the
samples used to construct the composite spectra. 
It is also now recognized that complete samples of
AGNs (at least those which are not Compton thick) are most easily
obtained with X-ray observations. Barger et al.\ (2003) and
Fontanot et al.\ (2007) have used X-ray samples to obtain 
direct limits on very high-redshift ($z=3.5-6.5$) ionizing fluxes.

%
% FIGURE 1 (galex_ion,1.15)
% all figures were width=3.5in
%
\vskip 0.5cm
\begin{inlinefigure}
\psfig{figure=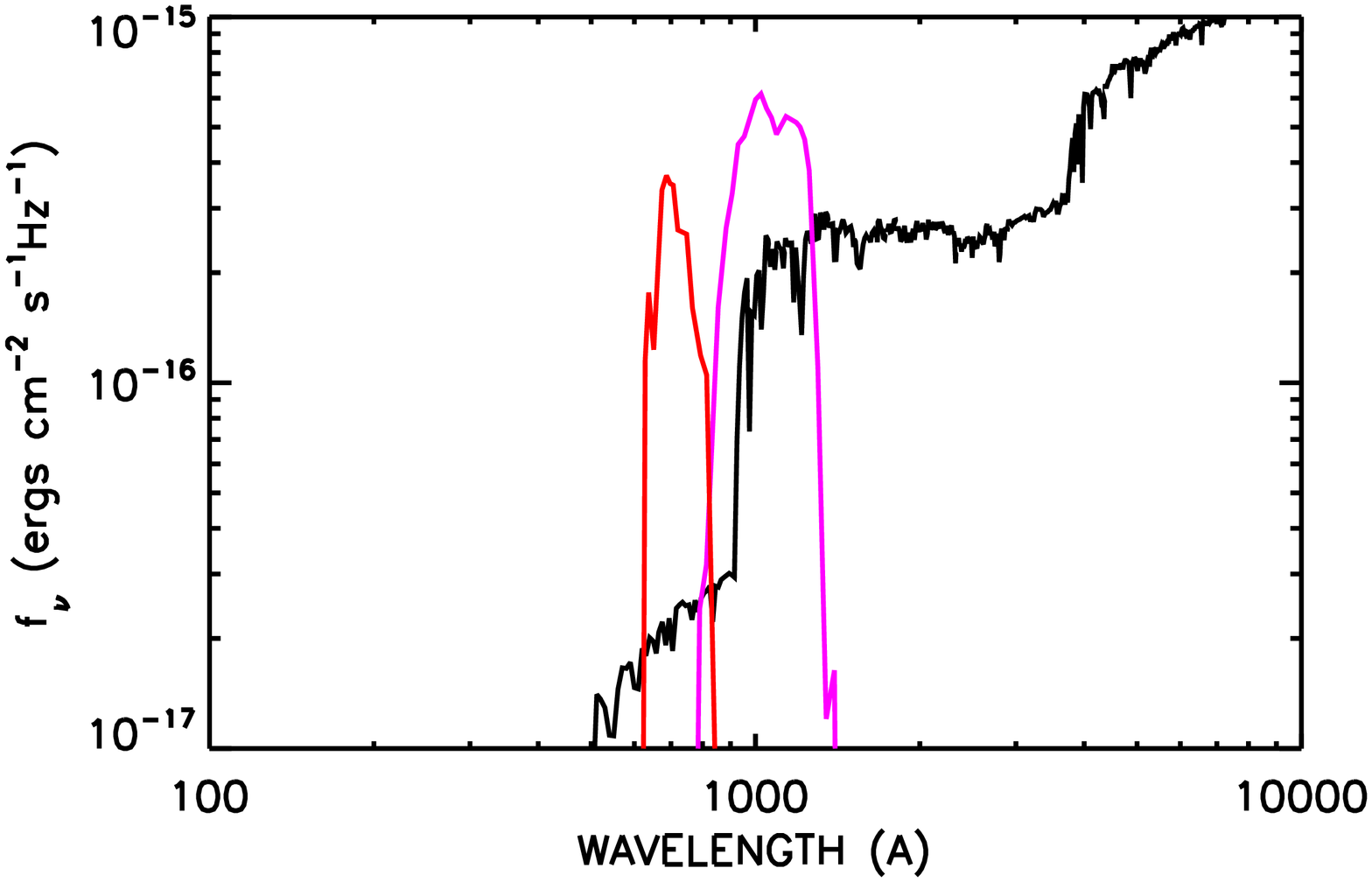,angle=90,width=3.5in}
\figurenum{1}
\figcaption[]{
The GALEX filter responses plotted for $z=1.15$ 
{\em (colored)\/} and the intrinsic spectrum of a galaxy 
with a decaying star formation rate {\em (black)\/} having
an exponentiation time of $5\times10^{9}$~yr and an age
equal to that of the universe at $z=1.15$. The galaxy
model is from Bruzual \& Charlot (2003) and does not
contain the effects of the galaxy's neutral hydrogen opacity
or internal extinction.
\label{filter_fig}
}
\end{inlinefigure}

In this paper we combine GALEX observations with X-ray samples to 
determine the ionizing flux from $z\sim1$ AGNs and then use this
result to estimate the AGN contribution to the metagalactic
ionizing background in the redshift range $z=0-5$. We 
also combine GALEX observations with a large optical galaxy sample with 
spectroscopic redshifts to determine limits on the escape of ionizing
photons at $z=1.15$. We then use this limit to constrain the galaxy 
contribution in the redshift range $z=0-5$. 
The structure of the paper is as follows. 
In \S\ref{secsample} we describe our optical and X-ray samples 
and how we determined the ultraviolet fluxes. In \S\ref{seccont}
we examine the contribution of optically selected galaxies and
X-ray selected AGNs to the ionizing flux. Only the X-ray selected
AGNs are positively detected.
In \S\ref{secxrayion} we determine how the ionizing flux relates
to other properties of the X-ray selected AGNs, showing that
the ionizing radiation is only seen in those with broad-lines
in their optical spectra. In \S\ref{secmet} we examine the 
evolution of the metagalactic ionizing 
flux. We show the ionizing fluxes as a function of redshift
and compare our results to previous calculations.
We summarize our results in \S\ref{secsummary}.

We assume $\Omega_M=0.3$, $\Omega_\Lambda=0.7$, and
$H_0=70$~km~s$^{-1}$~Mpc$^{-1}$ throughout.
All magnitudes are given in the AB magnitude system,
where an AB magnitude is defined by
$m_{AB}=-2.5\log f_\nu - 48.60$.
Here $f_\nu$ is the flux of the source in units of
ergs~cm$^{-2}$~s$^{-1}$~Hz$^{-1}$.

\section{Ultraviolet Fluxes for the Samples}
\label{secsample}

\subsection{Optical Sample}
\label{secopt}

We take as our initial optical sample all ${\rm F435W}<26$ galaxies in the 
145~arcmin$^2$ area of the ACS Great Observatories Origins Deep Survey-North 
(GOODS-N; Giavalisco et al.\ 2004) field. The  
F435W, F606W, F775W, and F850LP
magnitudes are taken from the ACS catalogs,
the $U$ magnitudes are from the $U$-band images of
Capak et al.\ (2004), 
and the spectroscopic redshifts are from Barger et al.\ (2008).
Further details and catalogs may be found in Barger et al. (2008).

The GALEX mission obtained a deep 150~ks exposure of
the GOODS-N region in early 2004. We measured the near-ultraviolet 
(NUV; 2371~\AA\ central wavelength) and far-ultraviolet
(FUV; 1528~\AA\ central wavelength) magnitudes at the positions of the 
optical sample using the GALEX
images, which we obtained from the Multimission Archive at STScI (MAST). 
Given the large point spread function of GALEX ($4\farcs5-6''$ FWHM), 
we used an $8'''$ diameter aperture to measure the magnitudes using the GALEX 
zeropoints of 20.08 for the NUV image and 18.82 for the FUV image 
from Morrissey et al.\ (2007).
We measured the magnitudes for the brighter objects (${\rm F435W}=20-23.5$) 
in both $8''$ and $24''$ diameter apertures and used the median offset 
of $-0.41$ between these measurements to
correct all of the $8''$ magnitudes to approximate total magnitudes.
The offset is slightly larger than the $-0.36$ correction from
$7.6''$ diameter magnitudes to total magnitudes determined by 
Morrissey et al.\ (2007). The magnitudes agree on average to within
0.05~mag with the SExtractor (Bertin \& Arnouts 1996) magnitudes given
in the GALEX NUV$+$FUV merged catalog for the region in the GR4
data release.

We measured
the noise level in the images by randomly positioning apertures on
blank regions of the sky and measuring the dispersion. (The procedure
for selecting the blank regions is outlined in \S3.1). We
found $1\sigma$ limits of 26.8 in the NUV image and
27.4 in the FUV image. Of the 6458 galaxies in the F435W sample,
1017 have ${\rm FUV}<25.5$ and 2528 have ${\rm NUV}<25$
(with considerable overlap between the two).

However, because of the large point-spread function (PSF) of GALEX, 
contamination by neighbors is a serious problem. 
This can be dealt with statistically,
as we discuss in \S\ref{seccont}, but it is also of considerable 
interest to look at the properties of the individual galaxies.
We therefore generated a subsample of isolated galaxies where
we eliminated any source which was closer than $8''$ to
a source with a brighter GALEX magnitude in the  ${\rm NUV}$ band 
or where, based on a visual inspection, the position was clearly
contaminated by the wings of a nearby bright GALEX source.
A substantial fraction of the galaxies in the optical sample are eliminated
by this isolation requirement, reducing the effective area to 48~arcmin$^2$
based on the fractional number of sources rejected by the isolation criteria.
The isolated galaxy sample contains 966 sources with ${\rm NUV}<25$ and 
422 sources with ${\rm FUV}<25.5$. The spectroscopic
identifications of these sources are 96\% complete to ${\rm NUV}=24.5$ 
and 98\% complete to ${\rm FUV}=25$.

\subsection{X-ray Samples}
\label{secx}

We construct our X-ray samples from the $2-8$~keV sources in the 
{\em Chandra\/} Large Area Synoptic X-ray Survey (CLASXS) of 
Yang et al.\ (2004), in the {\em Chandra\/} Lockman Area North Survey 
(CLANS) of Trouille et al.\ (2008), and within $8'$ of the pointing 
center of the 2~Ms {\em Chandra\/} Deep Field-North (CDF-N) survey 
of Alexander et al.\ (2003). 
We take as our full X-ray sample the sources with 
$f_{2-8~{\rm keV}}>3.5\times10^{-16}$~ergs~cm$^{-2}$~s$^{-1}$
(note that only the CDF-N contributes below 
$7\times10^{-15}$~ergs~cm$^{-2}$~s$^{-1}$). 
We take as our bright X-ray sample the sources with 
$f_{2-8~{\rm keV}}>7\times10^{-15}$~ergs~cm$^{-2}$~s$^{-1}$.
Current tables of redshifts for the full CLASXS, CLANS, and CDF-N 
X-ray samples are given in Trouille et al.\ (2008). 
We define  X-ray AGNs within the sample
on energetic grounds as any source more luminous than 
$L_X=10^{42}$~ergs~s$^{-1}$ (Zezas et al.\ 1998;
Moran et al.\ 1999). Here $L_X$ is the luminosity
calculated in the rest-frame $2-8$~keV band.  

The full X-ray sample contains 662 X-ray sources. Almost all of the sources
have been spectroscopically observed and 442 have spectroscopic redshifts.
The spectroscopically identified sources should contain a nearly complete
sample of all optical broad-line AGNs in the region (Barger et al.\ 2005;
Richards et al.\ 2005).
We find 171 broad-line AGNs, of which 98 have X-ray quasar luminosities
($L_X>10^{44}$~ergs~s$^{-1}$). The bright X-ray sample with
$f_{2-8~{\rm keV}}>7\times10^{-15}$~ergs~cm$^{-2}$~s$^{-1}$ contains 
485 sources, of which 390 have spectroscopic redshifts. 
Here we have 162 broad-line AGNs, of which 97 have X-ray quasar 
luminosities.

The CLANS and CLASXS fields are covered by a number of GALEX pointings,
all of which are roughly 30~ks in depth, somewhat shorter than the 
150~ks CDF-N GALEX exposure. The 1$\sigma$ limits are slightly
variable but typically about $26.1$
in the NUV images and $26.5$ in the FUV images.
Only a small fraction (about 4\%)
of the X-ray sources in the CLANS or CLASXS fields are not covered
by one of the  GALEX pointings and all of the CDF-N sources are covered.
The effective area for the X-ray sources
lying within the GALEX pointings is 0.9~deg$^2$
at $f_{2-8~{\rm keV}}$ above $7\times10^{-15}$~ergs~cm$^{-2}$~s$^{-1}$ and 
0.055~deg$^2$ below.

We  measured the GALEX magnitudes 
for the sources in the X-ray samples which lay within the
GALEX pointings in the same way 
that we measured the GALEX magnitudes for the GOODS-N optical sample.
Because the X-ray sources are generally brighter
in the FUV than the optical sources, 
the problem of contamination by galaxies is smaller, and we did not eliminate 
sources using a separation criterion. (The X-ray sources themselves
are so sparse that contamination by another X-ray source may be neglected.)
However, we did eliminate four
sources where visual inspection showed that there was contamination
from a nearby brighter GALEX source at the same wavelength. 
None of these four sources appear to have any significant FUV flux centered
on the object position.

\subsection{Redshift-Magnitude Relations}
\label{redmag}

In Figure~\ref{figz_fuv} we show redshift versus FUV magnitude 
for (a) the isolated optical sample with ${\rm FUV}<25$ and (b) the 
bright X-ray sample with ${\rm FUV}<25$. In both cases nearly all 
of the sources are spectroscopically identified, as can be seen from
the incompleteness histograms shown at the bottom of each panel. 
We have redshifts for 323 of the 329 sources with ${\rm FUV}<25$ 
in the optical sample. Five of the remaining sources could not be 
identified, and the final source has not been observed. All of the 
164 sources with ${\rm FUV}<25$ in the bright X-ray sample have been 
observed, and only 5 are not identified. In both panels we denote 
X-ray AGNs by red solid squares, and we enclose X-ray quasars in 
red large open squares.

At $z=0.6$ {\em (dashed horizontal lines in Figure~\ref{figz_fuv})\/}
the FUV filter straddles the Lyman break and substantially samples the ionizing 
radiation (see Fig.~\ref{filter_fig}). 
As we move to higher redshifts  nearly all galaxies vanish from the sample.
We can see from Figure~\ref{figz_fuv}a that the only source at $z>0.7$ 
in the optical sample with a significant ionizing flux is a broad-line
quasar. There is one additional galaxy at $z>0.7$, but the FUV
flux in this object is likely from a nearby $z=0.316$ galaxy,
which lies $2.7''$ away. (This source was not eliminated
from the sample because the companion is fainter in the NUV band.)

%
% FIGURE 2 (opt_z_fmag and xray_z_fmag)
%
\begin{inlinefigure}
\centerline{\psfig{figure=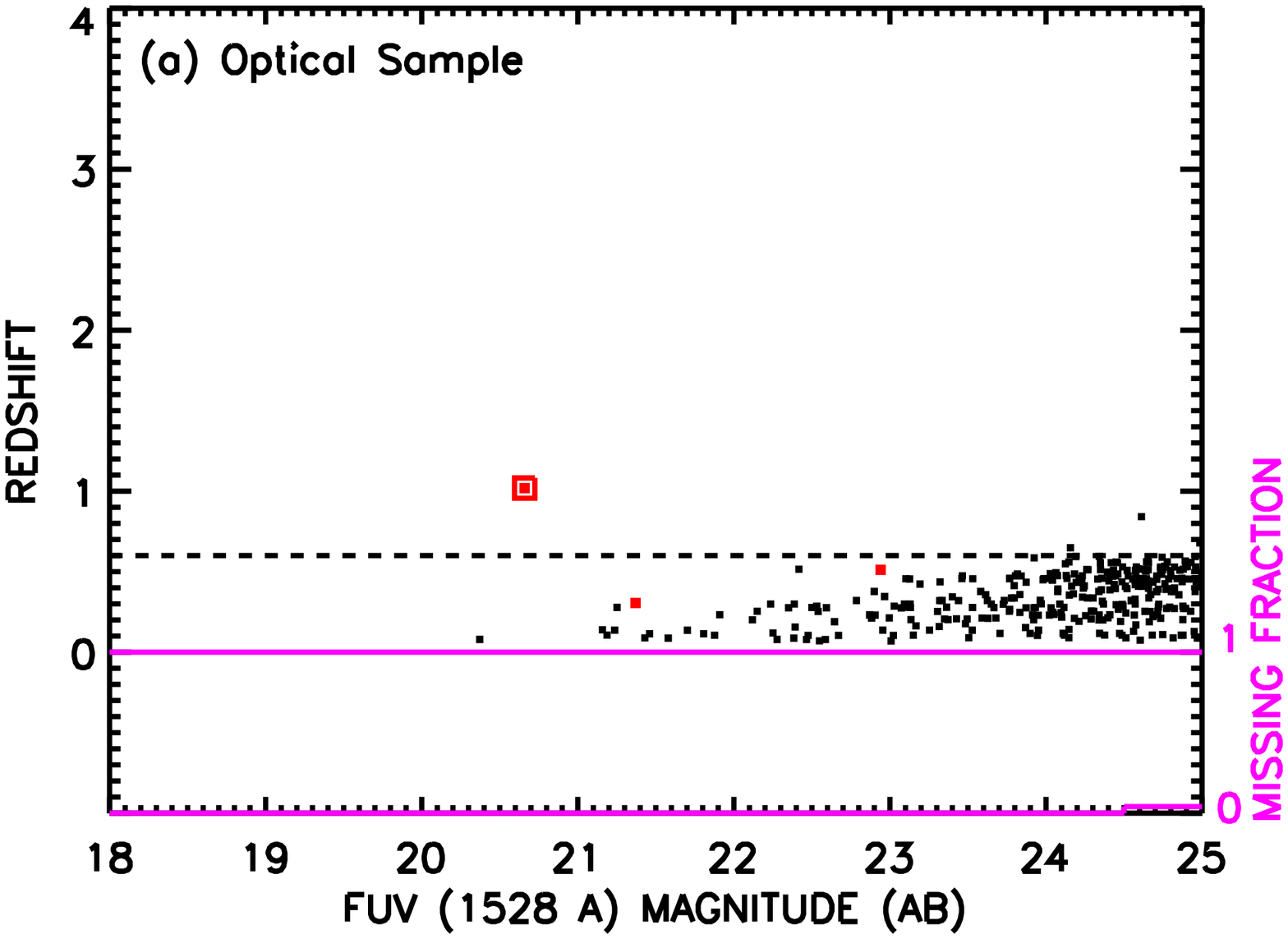,angle=90,width=4.0in}}
\vskip -0.5cm
\centerline{\psfig{figure=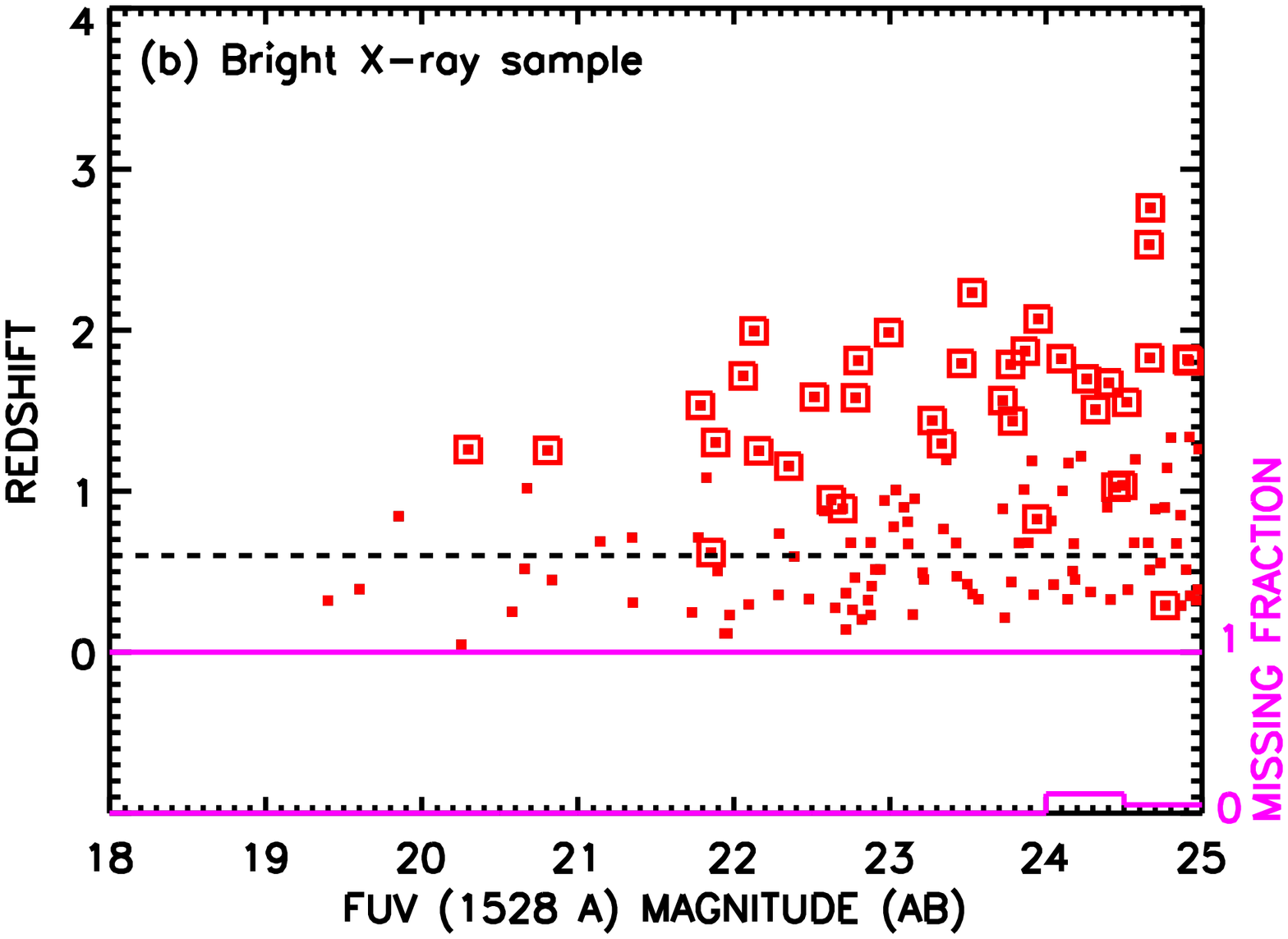,angle=90,width=4.0in}}
\figurenum{2}
\figcaption[]{
Redshift vs. FUV magnitude for (a) the isolated
optical sample (GOODS-N) and (b) the bright X-ray sample
(X-ray sources in the CLANS, CLASXS, and CDF-N fields with 
$f_{2-8~{\rm keV}}>7\times10^{-15}$~ergs~cm$^{-2}$~s$^{-1}$).
Sources with $L_X>10^{42}$~ergs~s$^{-1}$ (X-ray AGNs)
are denoted by red solid squares, and
those with $L_X>10^{44}$~ergs~s$^{-1}$ (X-ray quasars)
are shown enclosed in red large open squares. The fraction of
spectroscopically unidentified sources per
0.5~mag bin is shown in histogram form 
at the base of each panel. The $z=0.6$ redshift at which 
the FUV filter straddles the Lyman break is shown by a
dashed horizontal line in each panel.
\label{figz_fuv}
}
\end{inlinefigure}

We show this in a different way in Figure~\ref{fuv_redshift_hist},
where we plot the redshift distributions of all the spectrosopically identified
sources in the ${\rm F435W}<24.5$ GOODS-N sample
{\em (black histogram)\/} and of only those detected at ${\rm FUV}<25.5$ 
{\em (blue shading)\/}. (The GOODS-N 
sample is nearly spectroscopically complete to the F435W=$24.5$ limit.) 
We also plot the redshift distributions of all the X-ray AGNs present
in the optically selected ${\rm F435W}<24.5$ GOODS-N sample
{\em (green histogram)\/} and of only those detected at ${\rm FUV}<25.5$
{\em (red shading)\/}. Roughly 76\% of the galaxies in this sample
below $z=0.6$ have 
${\rm FUV}<25.5$. Of the 378 sources in this sample
above $z=0.7$, only four galaxies and two AGNs, including the one galaxy 
and the one broad-line quasar which we previously noted, have ${\rm FUV}<25.5$.

%
% FIGURE 3 (fuv_redshift_hist)
%
\begin{inlinefigure}
\centerline{\psfig{figure=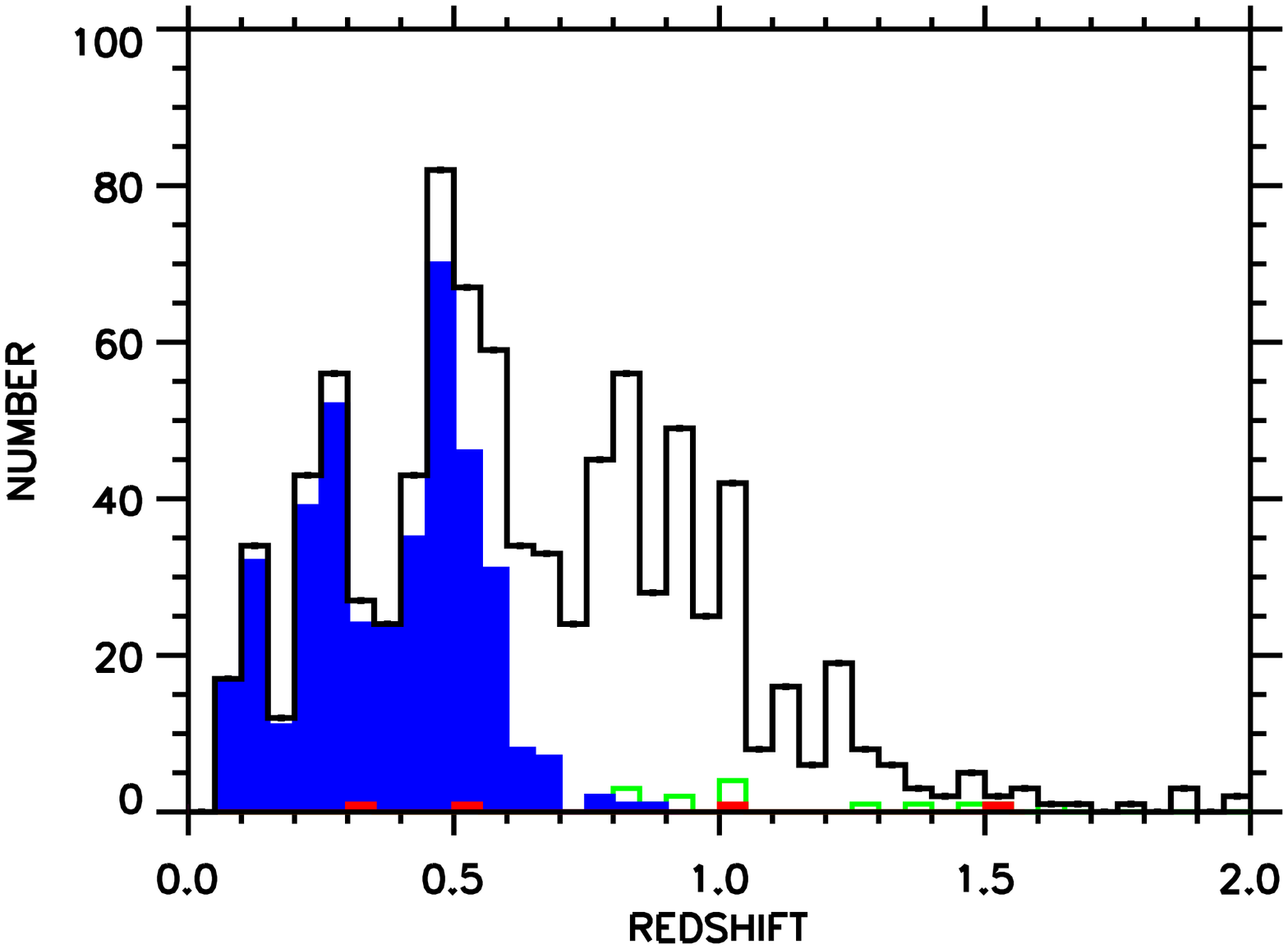,angle=90,width=4.0in}}
\figurenum{3}
\figcaption[]{The black histogram shows the spectroscopic redshift
distribution of the ${\rm F435W}<24.5$ sample in
the GOODS-N field. The binning interval is $0.05$ in redshift.
Galaxies which are detected with FUV magnitudes
less than 25.5 are shown with the blue shaded histogram.
The green histogram shows the spectroscopic redshift distribution of the
X-ray AGNs in this sample. X-ray AGNs which are detected with
FUV magnitudes less than 25.5 are shown with the red shaded
histogram.
\label{fuv_redshift_hist}
} 
\end{inlinefigure}

%
% FIGURE 4 (bl_quasars)
%
\vskip 0.5cm
\begin{inlinefigure}
\centerline{\psfig{figure=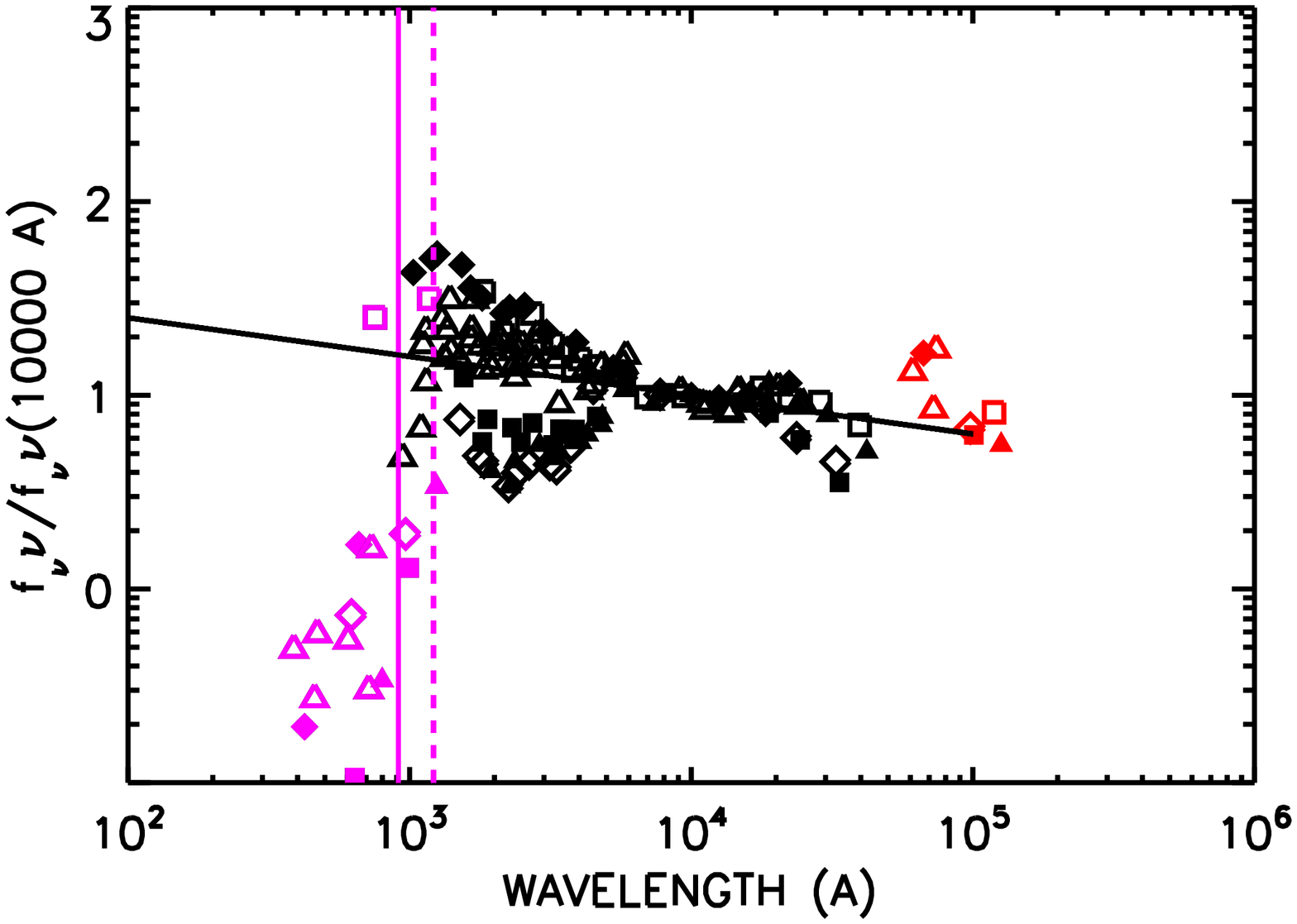,angle=90,width=3.5in}}
\figurenum{4}
\figcaption[]{
Spectral energy distributions of the seven broad-line 
quasars with $L_X>10^{44}$~ergs~s$^{-1}$ in the GOODS-N region having 
redshifts in the range $z=0.8-3$. Each quasar is shown with a unique 
symbol. The GALEX points are shown in purple. The 24~$\mu$m fluxes 
are from Treister et al.\ (2006) and are shown in red.
The purple solid line shows the Lyman break, and the purple dashed
line shows the position of Ly$\alpha$. The black solid line shows an 
$f_{\nu}$ vs. $\nu^{-0.8}$ power law. Only one source has a significant
ionizing flux {\em (open squares)\/}, and it appears to have almost 
no absorption.
\label{bl_quasars}
}
\end{inlinefigure}

The FUV-bright broad-line quasar is only one of a number of broad-line
quasars in the GOODS-N region, and it is immediately clear that these
quasars are extremely varied in the number of ionizing photons 
that we see from them. In Figure~\ref{bl_quasars} we show the spectral
energy distributions (SEDs) of the seven broad-line quasars with redshifts 
between $z=0.8$ and $z=3$ in the GOODS-N region. Nearly all of the quasars 
have significant breaks across the ionization edge, while the one quasar 
where we see a significant ionizing flux {\em (open squares)\/} shows a smooth 
extension from its longer wavelength SED and almost no opacity to the 
ionizing photons. A fraction of the breaks will be produced by intervening
Lyman Limit Systems (LLS). We quantify this in \S4.

The bright X-ray sample in Figure~\ref{figz_fuv}b, with its much larger 
area than the GOODS-N optical sample, contains a significant number of 
sources with detected ionizing fluxes that we can use to analyze the distribution 
in properties of the AGN ionizers and their contribution to the ionizing 
background.

\section{Contributions of Optically Selected Galaxies and 
X-ray Selected AGNs}
\label{seccont}

%
% FIGURE 5 (ion_hist and xray_ion_hist)
%
\vskip 0.2cm
\begin{inlinefigure}
\centerline{\psfig{figure=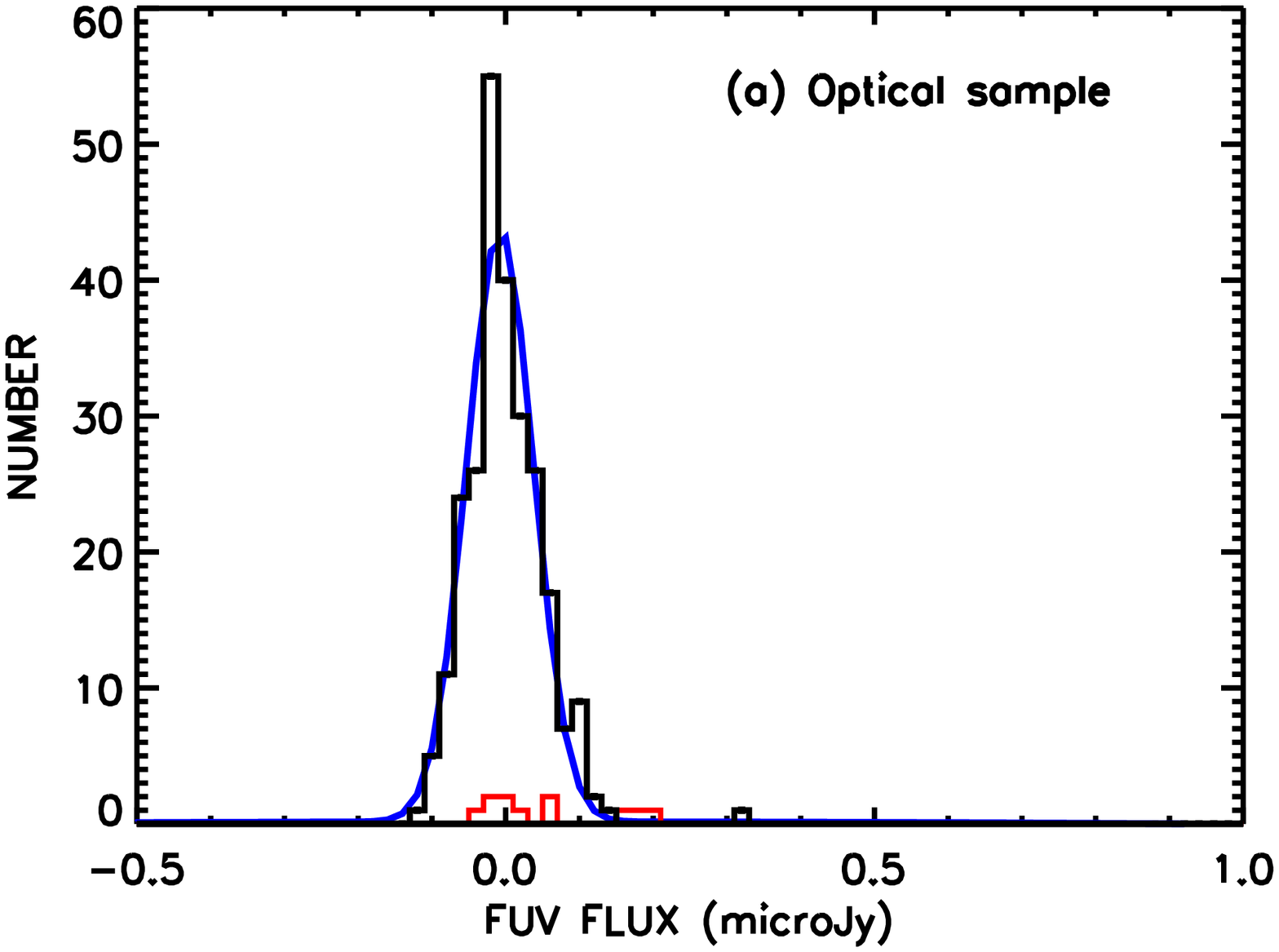,angle=90,width=3.5in}}
\centerline{\psfig{figure=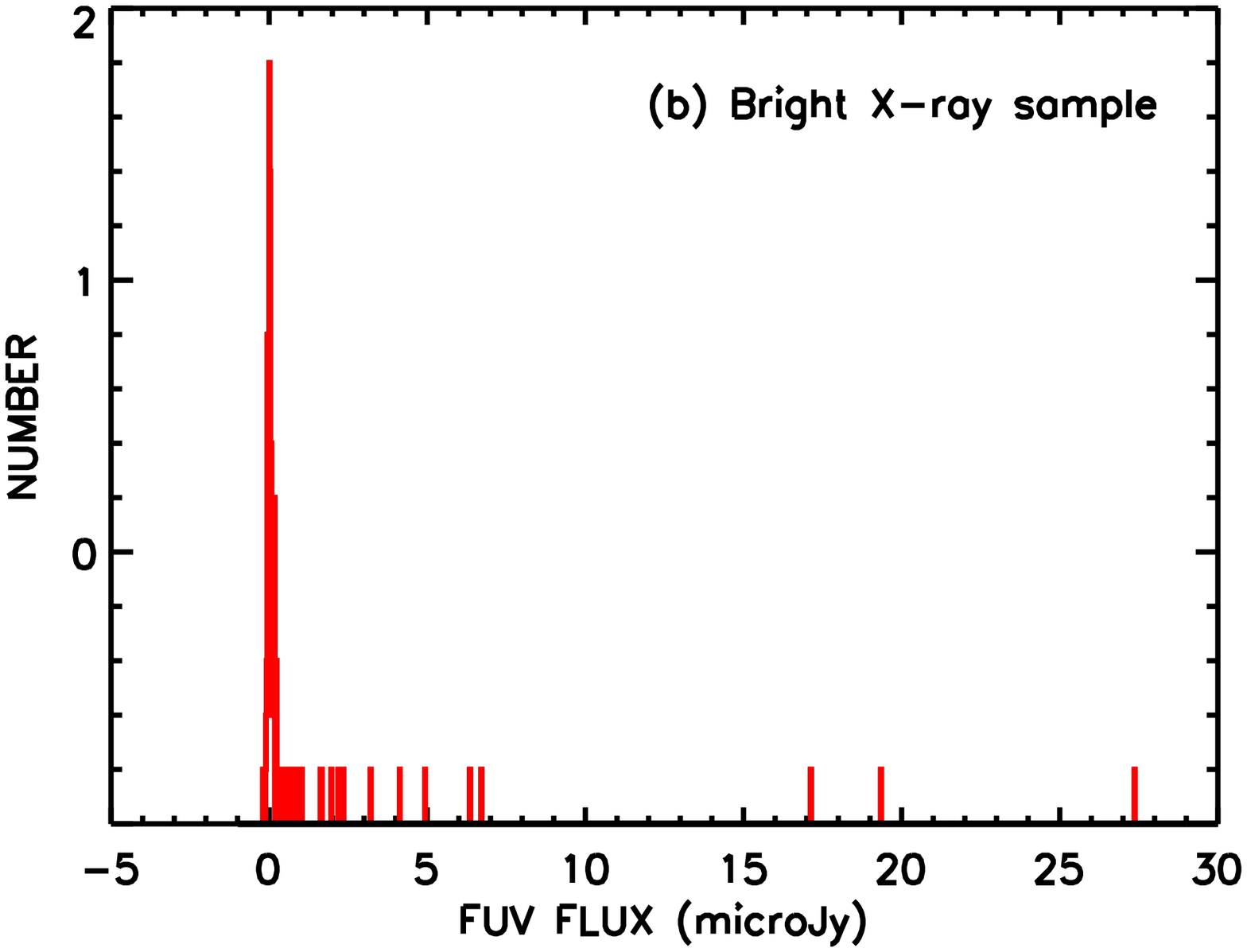,angle=90,width=3.5in}}
\figurenum{5}
\figcaption[]{
(a) Distribution of FUV fluxes in $\mu$Jy for the
isolated optical sample in the GOODS-N with $z=0.9-1.4$
(weak or no X-ray emission - {\em black histogram\/};
AGNs - {\em red histogram\/}). The smooth blue curve
shows the distribution of fluxes measured at random
positions in the FUV image.
All but the one broad-line quasar, which has a 19~$\mu$Jy flux,
are shown in the plot. 
(b) Same distribution for our bright X-ray sample with $z=0.9-1.4$
{\em (red histogram)\/}.
Note the much wider x-axis.
\label{ion_hist}
}
\end{inlinefigure}

In Figure~\ref{ion_hist}a we show the FUV flux distribution
of the isolated optical sample in the GOODS-N with 
$z=0.9-1.4$. In Figure~\ref{ion_hist}b we show the FUV flux distribution
of the bright X-ray sample with $z=0.9-1.4$.
The lower redshift cut-off of $z=0.9$ places the FUV filter 
well below the Lyman break. We have chosen a conservative upper 
limit on the redshift of $z=1.4$ to place the filter well above the 
drop-off in the intrinsic galaxy spectrum at about 500~\AA\ 
(see Fig.~\ref{filter_fig}) and to minimize the effects
of intervening intergalactic absorption.

\subsection{Optical Galaxy Sample}
\label{secoptsam}

In Figure~\ref{ion_hist}a we show the FUV flux distributions of the sources
with weak or no X-ray emission {\em (black histogram)\/} and of the X-ray
AGNs {\em (red histogram)\/} in the isolated optical sample in the GOODS-N 
with $z=0.9-1.4$. We overplot the distribution of fluxes measured at random 
isolated positions in the GOODS-N FUV image {\em (blue curve)\/}, which
matches extremely well to the black histogram. The one
detected galaxy in Figure~\ref{figz_fuv}a is seen at 
0.3~$\mu$Jy, but the one detected broad-line quasar
is far off the scale of the plot.
Excluding only the broad-line quasar while keeping the other
X-ray AGNs, these isolated galaxies formally
contain a total positive signal of $3.0\pm0.8~\mu$Jy. However, when
we assess this result we must be concerned about contamination 
by foreground galaxies and systematic effects of the 
background subtraction. Indeed, the average FUV flux per galaxy 
is only 0.011~$\mu$Jy, which is considerably below the
$1\sigma$ flux of 0.039~$\mu$Jy in the image.

These effects are best tested by Monte-Carlo simulations using random 
position samples. This also allows us to avoid any subjectivity that might be 
present in the isolation criteria, since we can analyze analytically selected 
samples and treat the effects of blending using the simulations. The present 
problem is also simplified since the contaminating effects primarily arise from 
the $z \lesssim 0.6-0.7$ galaxies which, as we have seen in Figure~\ref{fuv_redshift_hist}, 
are by far the most common class of source with substantial FUV emission. 
Since we do not expect these to be correlated with the $z\sim1$ galaxies of 
interest, we do not need to be concerned with the effects that might
be introduced by correlations, and we can use spatially uniform random 
samples to determine our backgrounds.

The simplest procedure is to create an analytically selected isolated galaxy 
sample. To do this we masked $16''$ radius regions around 
all $z=0-0.7$ galaxies and stars with ${\rm FUV}<23$ and $8''$ 
radius regions around all $z=0-0.7$ galaxies 
and stars with ${\rm FUV}<25.5$ and excluded all $z=0.9-1.4$
galaxies which lay in the masked regions from our sample.
In total, we masked regions
around 469 galaxies and 14 stars. We also masked $16''$ radius
regions around the four X-ray AGNs in the field with ${\rm FUV}<25.5$. 
These isolation criteria removed 187 galaxies 
from our initial sample of 680 galaxies with $z=0.9-1.4$ in the GOODS-N region
and reduced the effective area to 104~arcmin$^2$. The average FUV flux per 
galaxy in the analytic isolated sample is 0.012~$\mu$Jy, 
which is similar to what
we found for the smaller isolated sample we constructed in \S\ref{secopt}.
  
We next created 100 random realizations, each containing
the same number of positions as the target sample. 
We constrained the random positions to lie away from the low-redshift
galaxies, stars, and AGNs in the same way that we constrained the target
sample. However, since the signal associated
with the $z=0.9-1.4$ galaxies is small, we did not remove or
avoid the actual sample when constructing the random samples,
but simply generated random $x$ and $y$ positions within the 
unmasked area. From the 100 random realizations we found that a random
position satisfying the isolation criteria has an average
FUV flux of 0.018~$\mu$Jy and a dispersion of 0.003~$\mu$Jy. Subtracting
this from the average FUV flux per galaxy of 0.012~$\mu$Jy measured
from the analytic isolated target sample 
gives an FUV flux of $-0.006\pm0.003~\mu$Jy per galaxy.

The average NUV flux per galaxy in the same sample
is $0.44~\mu$Jy, while the average $U$-band flux per galaxy
is $0.96~\mu$Jy. (The corresponding fluxes for a random
position at these wavelengths are negligible compared to these values.)
The rest-frame wavelengths of the two
bands at the midpoint of the $z=0.9-1.4$ redshift interval are 1100~\AA\
and 1720~\AA, respectively. The NUV band straddles the Lyman
continuum break in the higher redshift portion of the redshift interval 
which will reduce the observed flux in this bandpass and it
may also be reduced by the Lyman alpha forest (see Fig.~\ref{filter_fig}). 
We therefore chose to use the $U$-band flux as a measure of the rest-frame flux 
at 1500~\AA, assuming a flat $f_\nu$ SED. We then formed the ratio
of the rest-frame 700~\AA\ flux to the rest-frame 1500~\AA\
flux, ignoring the small differential $K$-correction as a function
of redshift. We call this ratio the ionization fraction to distinguish
it from the escape fraction of ionizing photons from the galaxy.
Conversion of the ionization fraction to the escape fraction
requires knowledge of the intrinsic galaxy spectrum and reddening
(e.g., Haardt et al.\ 1999; Siana et al.\ 2007). However, for
the present purposes of obtaining the ionizing emission of the
galaxy population, it is the ionization fraction which is
of interest and which we can use to convert the rest-frame
UV emission to the ionizing emission. The ionization fraction
we obtained using the background-corrected analytic isolated sample is
$-0.006\pm0.003$. 

An alternative way to proceed is to subtract the
foreground galaxies and stars to form a cleaned image
and then to measure the signal associated with the target
sample on the cleaned image. This has the virtue of allowing
us to measure a larger fraction of the target sample and to
avoid using the somewhat arbitrarily sized masking regions selected
in the previous procedure. 
To form a cleaned image we fitted and subtracted all of the $z<0.7$ galaxies 
with ${\rm FUV}<25.5$ in the GOODS-N from the GALEX image, as well as all 
of the stars and the four X-ray AGNs with ${\rm FUV}<25.5$. 
As can be seen from Figures~\ref{figz_fuv}a and \ref{fuv_redshift_hist}, 
this should remove nearly all of the signal from directly detected objects
and therefore minimize the effects of blending and optimize
the background subtraction.

Our procedure for this cleaning was to normalize the
GALEX PSF, which we determined by stacking the FUV detected
stars in the image, to the measured flux of each galaxy
and then to subtract this from the image. Most galaxies
are very compact compared to the broad GALEX PSF, but to remove
the 36 brighter objects with ${\rm FUV}<22.5$, which may be more extended,
we masked out a $12''$ region around these galaxies.
In our subsequent analysis we exclude objects which fall
within these masked regions and correct the area appropriately.

For each subset sample from the full ${\rm F435W}<26$ sample to be 
analyzed, we measured the fluxes in the cleaned image
at the actual positions of the sample and then 
created 100 random samples with an equal number of positions as 
the sample to determine the background and dispersion. We then 
subtracted this background from the measured signal of the 
sample to obtain the true signal. Again, since the signal associated
with the $z=0.9-1.4$ galaxies is small, we did not remove or
avoid the actual sample when constructing the random samples.
%but simply generated random $x$ and $y$ positions within the area.

For the 626 spectroscopically identified $z=0.9-1.4$ galaxies 
with ${\rm F435W}<26$ that are not X-ray AGNs and are
not in the small masked areas, we find an average background-corrected
FUV flux per galaxy of $-0.005\pm 0.002~\mu$Jy. 
The average NUV flux per galaxy is $0.51~\mu$Jy,
and the average $U$-band flux per galaxy is $0.97~\mu$Jy.
We note again that the NUV and U signal associated with
a random postion is negligible.
The inferred ionization fraction is $-0.006\pm0.003$.
This procedure gives a consistent result
with the analytic isolation simulations described above. 
Both give negative fluxes but are consistent with a zero flux
at just over the $2\sigma$ level.

We can also directly compare the distribution of fluxes in the observed 
target sample with that from the total of all 100 realizations 
of the random sample to test if they are different
from one another. In Figure~\ref{ks} we compare the cumulative
distribution of the FUV fluxes for the $z=0.9-1.4$ ${\rm F435W}<26$ sample 
measured from the cleaned image {\em (black curve)\/} with that measured
from the random sample {\em (red curve)\/}. 
A Kolmogorov-Smirnov test give a 1\% probability
that the target sample has the same distribution as the random sample.

%
% FIGURE 6 (ks)
%
\begin{inlinefigure}
\psfig{figure=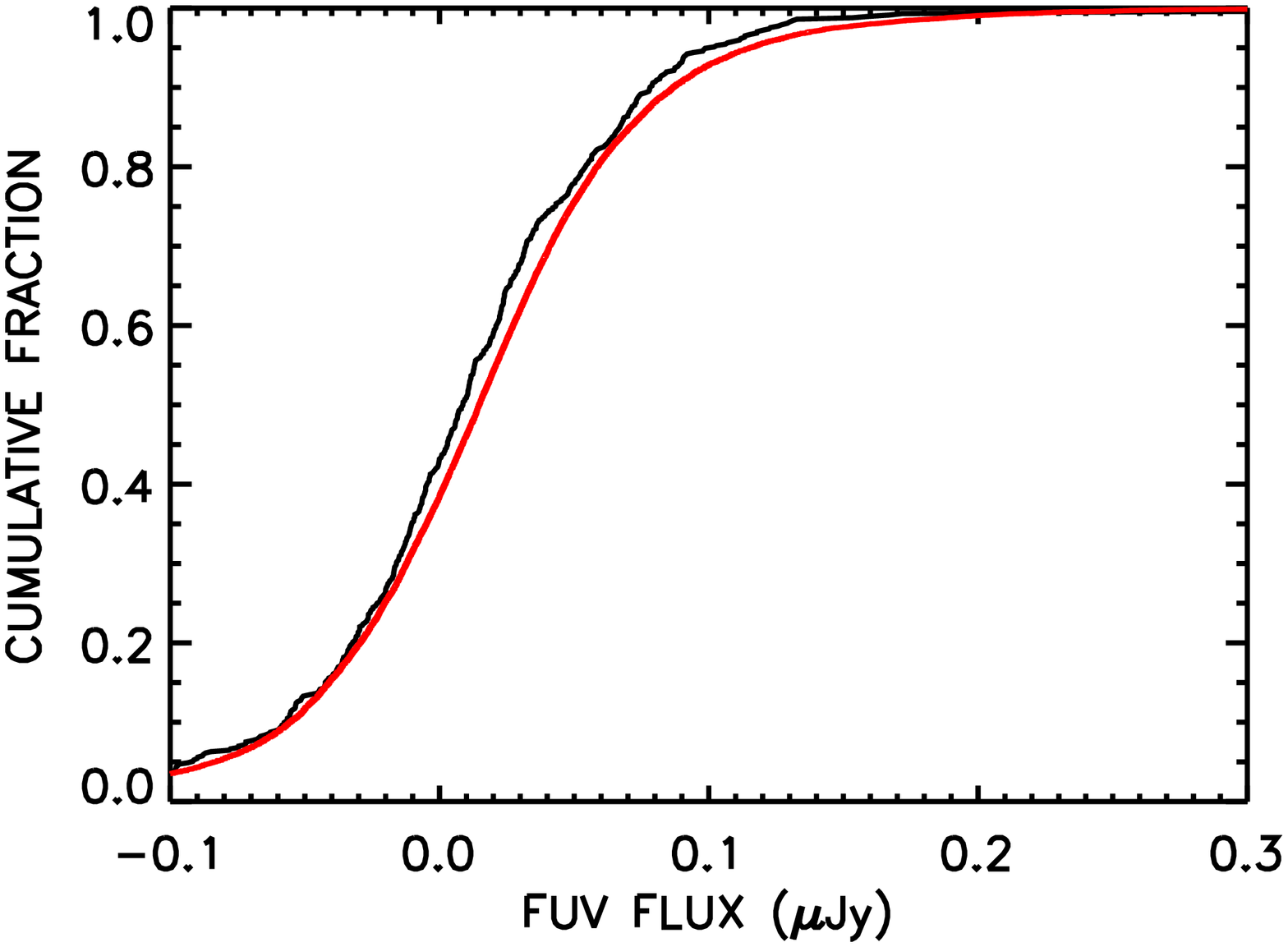,angle=90,width=4.0in}
\figurenum{6}
\figcaption[]{
Cumulative fraction of sources vs. FUV flux measured in the cleaned image. 
The black curve shows the ${\rm F435W}<26$ sources in the $z=0.9-1.4$
redshift interval, and the red curve shows the random sample.
\label{ks}
}
\end{inlinefigure}

For the nearly spectroscopically complete ${\rm F435W}<24.5$ sample, there 
are 363 $z=0.9-1.4$ galaxies that are not X-ray AGNs and are not in the
small masked areas. Here we find an average background-corrected
FUV flux per galaxy of $-0.005\pm 0.003~\mu$Jy. The average NUV flux per galaxy
is $0.65~\mu$Jy, and the average $U$-band flux per galaxy is $1.34~\mu$Jy.
The inferred ionization fraction is $-0.005\pm 0.003$.

We should probably not take the negative signal
too seriously, given its statistical significance 
in all of the measurements described above.
However, the average negative flux
may seem worrying at first sight and suggestive of some type
of systematic error in the procedure. We tested this by measuring
the signal for the 229 galaxies with $z>1.6$ with the analytic isolation
procedure. Here we find an average background-corrected FUV flux 
per galaxy of $0.001\pm 0.004~\mu$Jy.
Thus, the negative signal seems to be associated only with
the specific target sample.

However, it is worth noting that the galaxies could have an
effective negative signal if their shadowing effect
on the ionizing background exceeded their direct emission.
The galaxies will block any background light at these wavelengths
over the region where their neutral hydrogen column density makes them
a LLS.  The shadowing effect may extend over a larger
area than any FUV emission, making the larger apertures needed for the
GALEX observations more sensitive to
it. Such shadow galaxies could be used
to make a measure of the ionizing flux and of the extent
of the neutral hydrogen in the individual galaxy. However, higher
spatial resolution observations would be more appropriate for
such an analysis than the GALEX data.

In all of our estimates above,
systematics could raise the absolute value
of the error, though they will not change the ratio of the signal
to the error. In particular, we should allow for uncertainties
in the extrapolation to form the 1500~\AA\ flux, possible
systematic errors in the determination of the FUV fluxes
(Morrissey et al.\ 2007), and uncertainties in the aperture corrections
to total magnitudes. 
Specifically, we include the possibility of a systematic error of 0.2
in the relative FUV  and $U$-band magnitude determinations, which
would raise the error in the ionization fraction to 0.004. Allowing
for this level of systematic error and also
ignoring the negative signal, we take the $2\sigma$
limit of 0.008 as our upper limit on the ionization
fraction in the $z\sim1$ galaxy population in what follows.

Given the relatively small size of the GOODS-N field, we might
also be concerned that we are missing the rare FUV-bright
galaxies. Since we see no FUV-bright galaxies in the field, we
adopt a $1\sigma$ upper limit of 1.8 such galaxies (Gehrels 1986).
Adopting a typical value for the FUV magnitude of 23---comparable
to the brightest $z\sim0.6$ galaxies (Fig.~\ref{figz_fuv})---one
such bright galaxy would correspond to an average 
contribution of 0.006~$\mu$Jy per galaxy from the 680 measured galaxies. 
This is already smaller than the adopted limit. However, using an FUV magnitude 
of 23 is in fact extreme, since even young star-forming galaxies 
will have a strong break across the Lyman edge, so the true
contribution will be smaller. It does remain conceptually
possible that there may be unique environments where UV ionizing
galaxies preferentially form and that we simply have not sampled
these with the GOODS-N area. This can only be tested with larger
area samples.

In Figure~\ref{escape_fig}a we show the ratio of the rest-frame 700~\AA\ flux 
to the rest-frame 1500~\AA\ flux (the ionization fraction) as a function 
of the absolute
rest-frame 2000~\AA\ magnitude computed from observed-frame F435W {\em (black squares)\/}. 
Here we use the cleaned FUV images and do the background subtraction as described above.
The red line shows the $2\sigma$ upper limit of 0.008 on 
the average ionization fraction for the full ${\rm F435W}<26$ sample.
In Figure~\ref{escape_fig}b we show the ionization fraction
versus observed-frame ${\rm F435W}-{\rm F850LP}$ color (roughly
rest-frame 2000~\AA\ $-$ 4200~\AA).
There is no significant detection as a function of either luminosity or color.

%
% FIGURE 7 (escape)
%
\begin{inlinefigure}
\psfig{figure=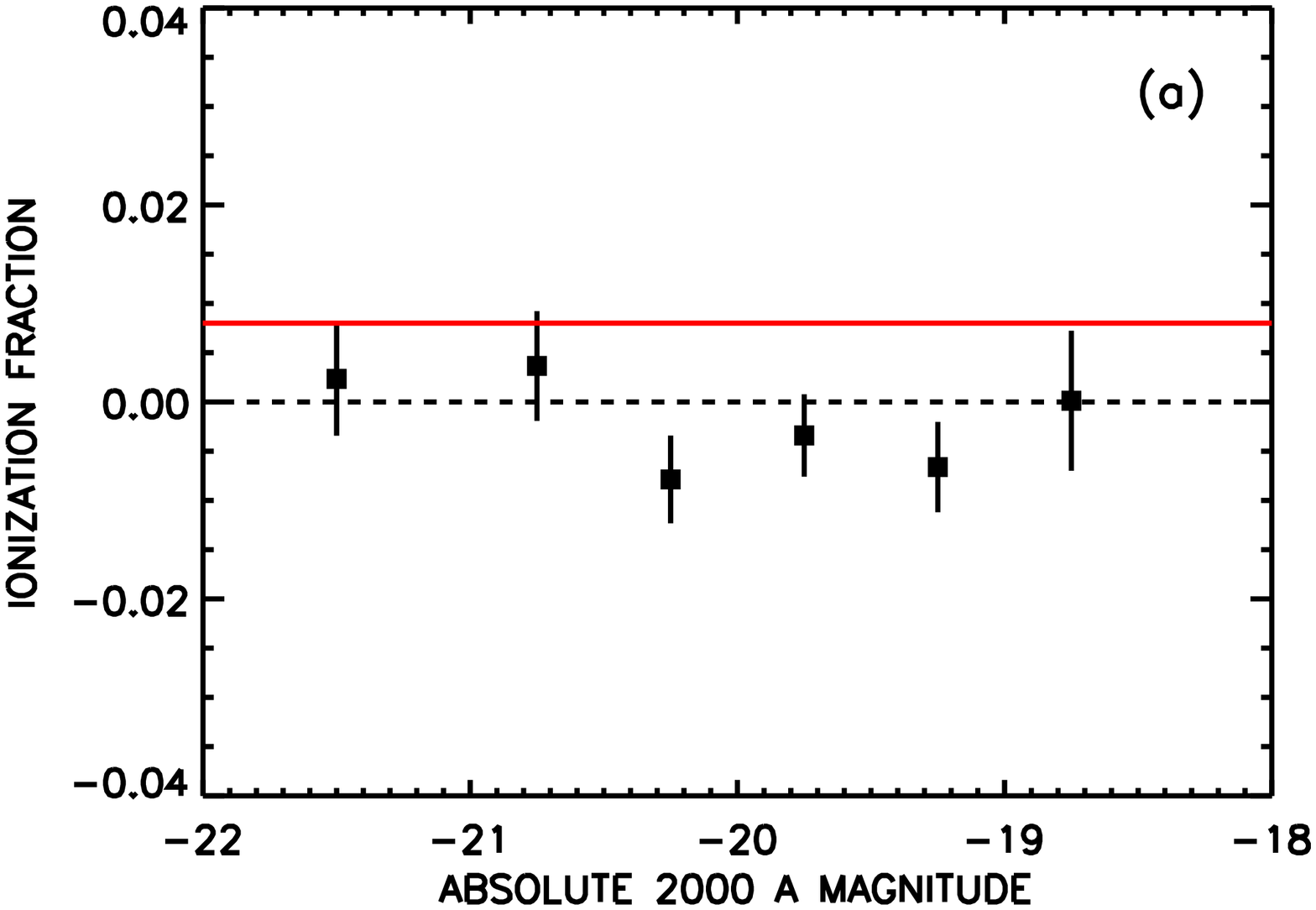,angle=90,width=4.0in}
\psfig{figure=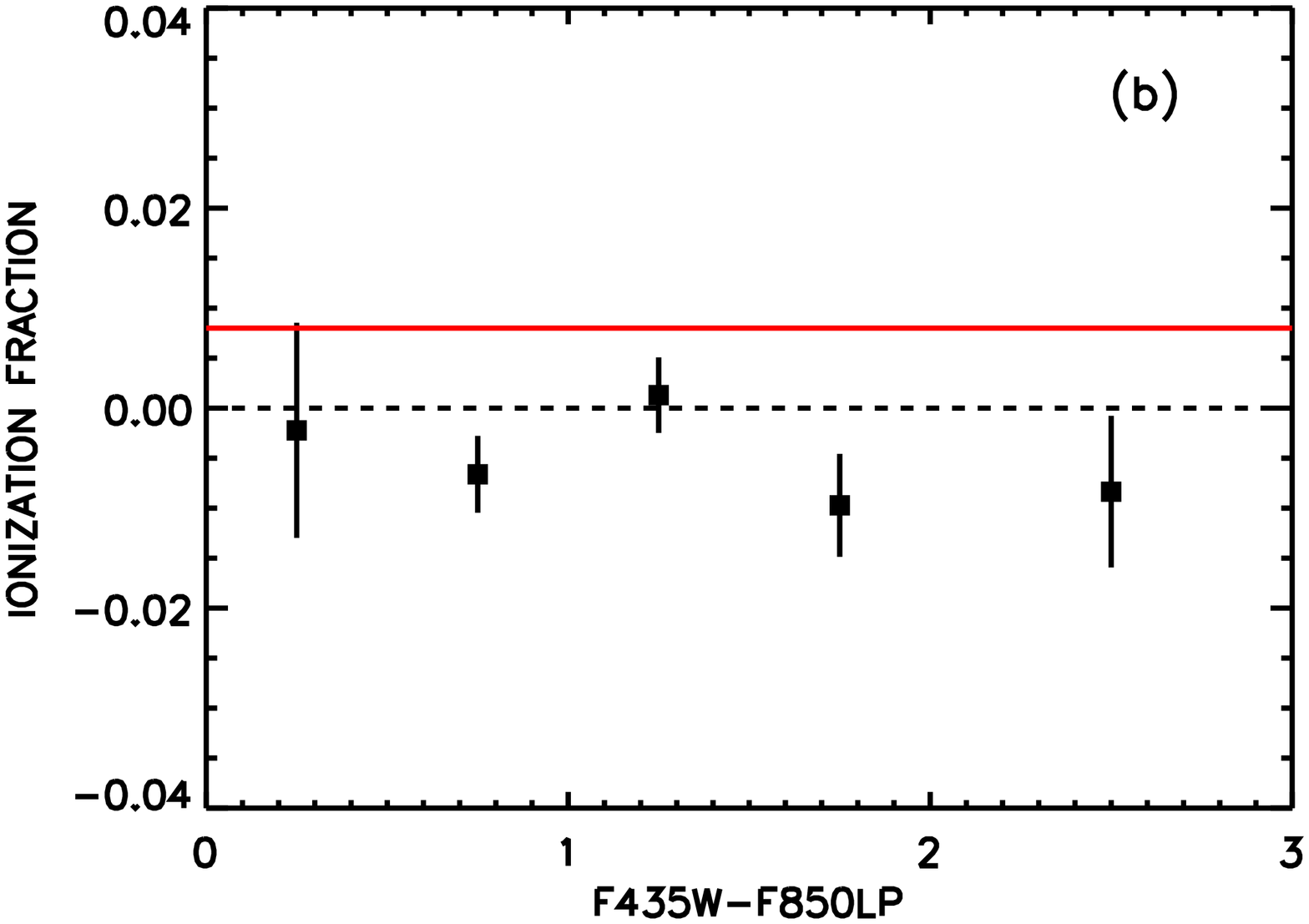,angle=90,width=4.0in}
\figurenum{7}
\figcaption[]{
(a) Average ionizing fraction (ratio of the rest-frame 700~\AA\ flux to the 
rest-frame 1500~\AA\ flux) vs. absolute rest-frame 2000~\AA\ magnitude computed 
from observed-frame F435W. The black squares show the results
for the spectroscopically identified $z=0.9-1.4$ galaxies
in the ${\rm F435W}<26$ sample that are not X-ray AGNs.
The error bars are $1\sigma$ statistical errors. The red solid line shows the
$2\sigma$ upper limit of 0.008 on the average ionization fraction for the full 
sample, including possible systematic errors described in the text.
(b) Average ionizing fraction vs. observed frame F435W$-$F850LP
color (roughly rest-frame 2000~\AA\ - 4200~\AA) for the $z=0.9-1.4$ galaxies. 
\label{escape_fig}
}
\end{inlinefigure}

\subsection{Bright X-ray Sample}

\begin{figure*}
\centerline{\psfig{figure=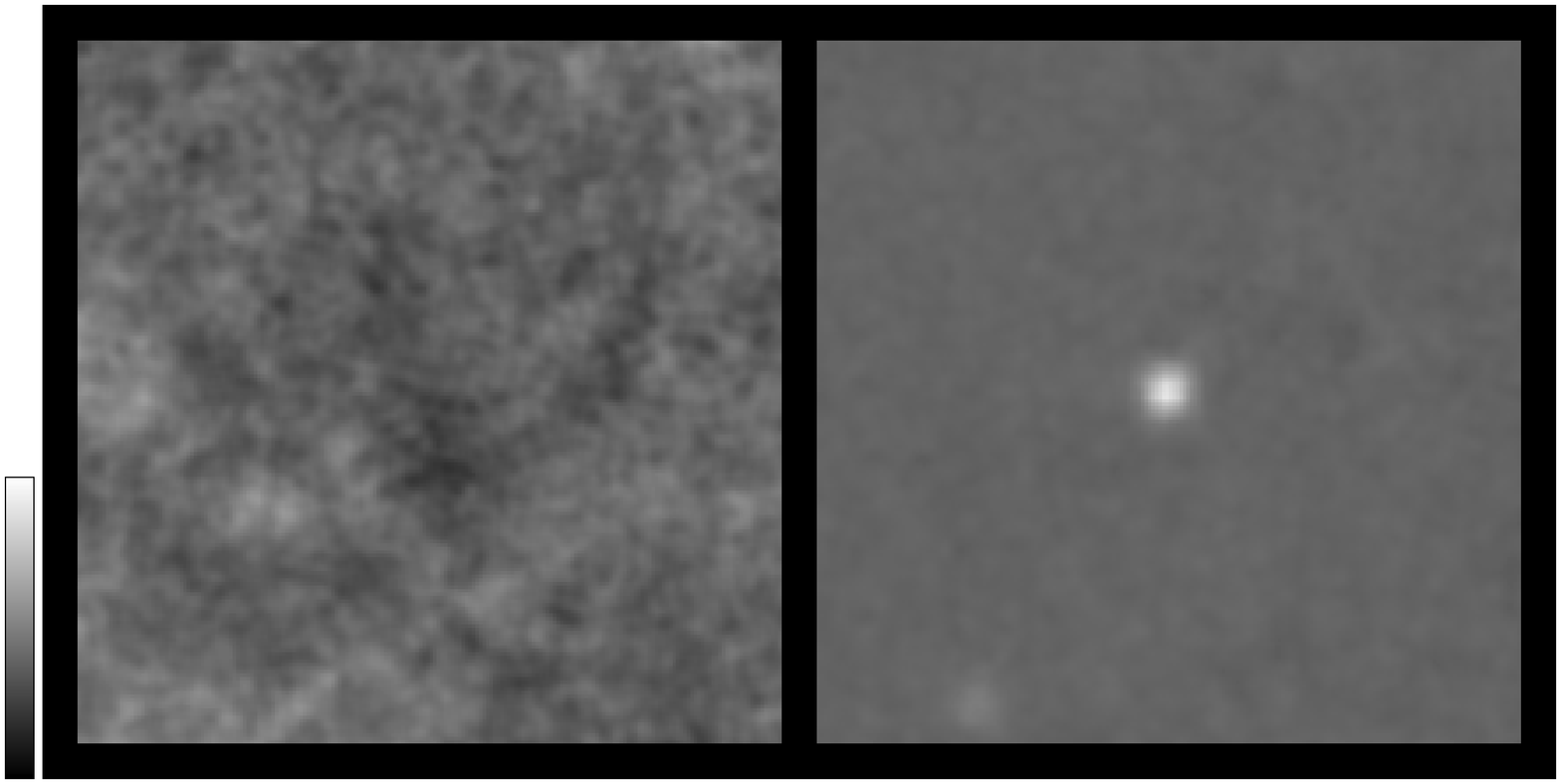,angle=90,width=3.5in}}
\figurenum{8}
\figcaption[]{
Summed image of the 626 spectroscopically
identified $z=0.9-1.4$ galaxies in the ${\rm F435W}<26$ sample that
are not X-ray AGNs and do not lie in the small number
of masked areas in the cleaned image
{\em (left panel)\/} and the 29 GOODS-N X-ray AGNs
with $z=0.9-1.4$ {\em (right panel)\/}.
The noise is higher in the galaxy image because of the
larger number of objects.
The scaling is the same in both images. In each case the image 
size is $80''$ on a side, and a positive signal would appear at 
the center of the image. The strong detection in the right panel 
is dominated by the one FUV-bright broad-line quasar. There is 
no detected signal in the left panel.
\label{stack_images}
}
\end{figure*}

The one FUV-bright broad-line quasar in the GOODS-N
gives a flux of 19~$\mu$Jy, which dominates the signal in the
field. We show this in Figure~\ref{stack_images}, where we 
compare the summed image formed using our cleaned image
of the 626 spectroscopically identified
$z=0.9-1.4$ galaxies in the ${\rm F435W}<26$
sample that are not X-ray AGNs {\em (left panel)\/}
with the summed image of the 29 GOODS-N X-ray AGNs 
with $z=0.9-1.4$ {\em (right panel)\/}.
The clear detection in the FUV sum of the X-ray AGNs is totally dominated 
by the FUV-bright broad-line quasar. The remaining 28 X-ray AGNs
in this redshift interval are consistent with having no ionizing flux.
However, given the small number of contributing broad-line AGNs,
we need the much wider field of the bright X-ray sample to 
accurately compute the AGN contribution to the ionizing flux.

%
% FIGURE 8 (stack_pic)
%

In Figure~\ref{ion_hist}b we show the FUV flux distribution for our bright X-ray 
sample with $z=0.9-1.4$ {\em (red histogram)\/}. 32 sources in this redshift 
interval have FUV magnitudes brighter than the $5\sigma$ limit of the image, 
and the total measured flux is $112~\mu$Jy for the 0.9~deg$^2$ area. 
This result is not affected by our exclusion of the four sources which are 
contaminated by neighbor galaxies (see \S\ref{secx}), since,
even without any cleaning, these sources only have a total flux of $2~\mu$Jy.
We analyzed the background level and variance 
by again creating random realizations, each containing the same number of
positions as the target sample, and measuring at the positions in the raw images. 
We found the background to be small relative to the signal.
The background-corrected signal is $100\pm5~\mu$Jy, where the error is 
the variance in the background determination. This measured 
signal would correspond to 4.5~$\mu$Jy in the small GOODS-N 
region. Thus, the one FUV-bright broad-line quasar in the 
GOODS-N region is an anomaly. 

Even with the larger area, there are 
only a small number of AGNs which are seen to have ionizing
radiation. The three brightest AGNs produce 60\% of the flux, and 
the seven brightest AGNs produce almost 80\%. Given the small number 
of contributing sources and the peculiar distribution of the fluxes, 
we have used a jackknife analysis to make a more accurate estimate 
of the uncertainty in the AGN contribution.
This gives a 68\% confidence range of $\pm33~\mu$Jy on the signal.
It should be noted that large uncertainty in the jackknife analysis 
does not mean that the  ionizing flux contribution from the AGN is not
highly significant. This is properly inferred from the results
in the last paragraph.
The large uncertainty from the jackknife method is a consequence of the 
fact that only a small fraction of the X-ray sources which are
detected in the FUV dominate the production of ionizing radiation. 

We conclude that only the X-ray sources
can be detected at ionizing frequencies and that
an extremely small number of these dominate the ionizing
flux production. The emission per unit comoving volume of the 
X-ray quasars ($L_X>10^{44}$~ergs~s$^{-1}$) is 
$\nu \lambda_{\nu}=(3.4\pm1.5)\times10^{39}$~ergs~s$^{-1}$~Mpc$^{-3}$ 
at $z=1.15$, where we have used the jackknife method 
to estimate the uncertainty. Here $\lambda_{\nu}$ is the
luminosity density per unit frequency, which is the sum of the luminosities 
per unit frequency of
all the objects divided by the cosmological volume that they occupy.
We now turn to a more detailed analysis of the contribution 
from the bright X-ray sample to try to understand the FUV-bright
AGNs in more detail.

\section{Ionizing Properties of X-ray Selected AGNs}
\label{secxrayion}

\begin{figure*}
\figurenum{9}
\epsscale{1}
\centerline{\psfig{figure=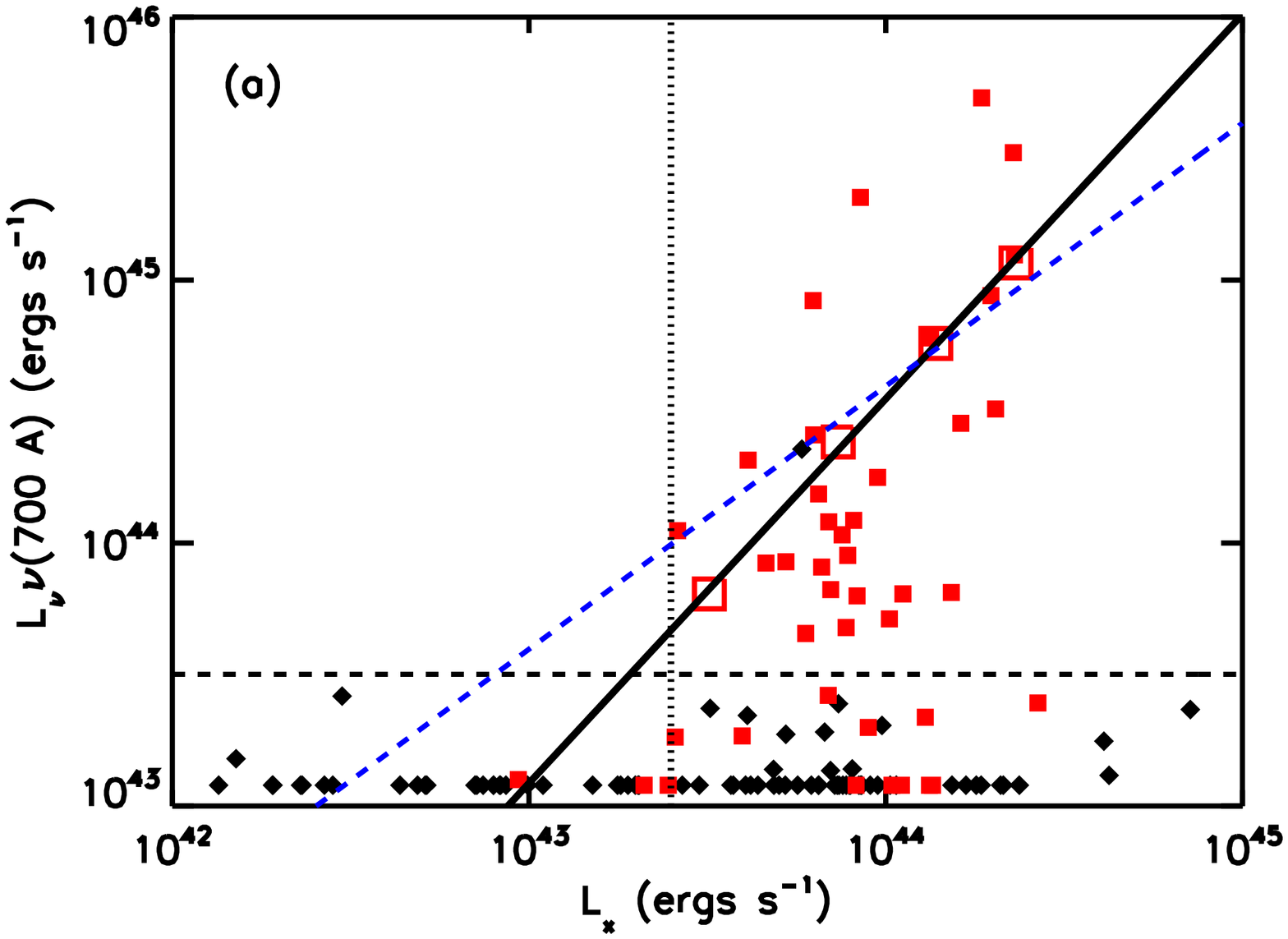,angle=90,width=3.8in}}
\centerline{\psfig{figure=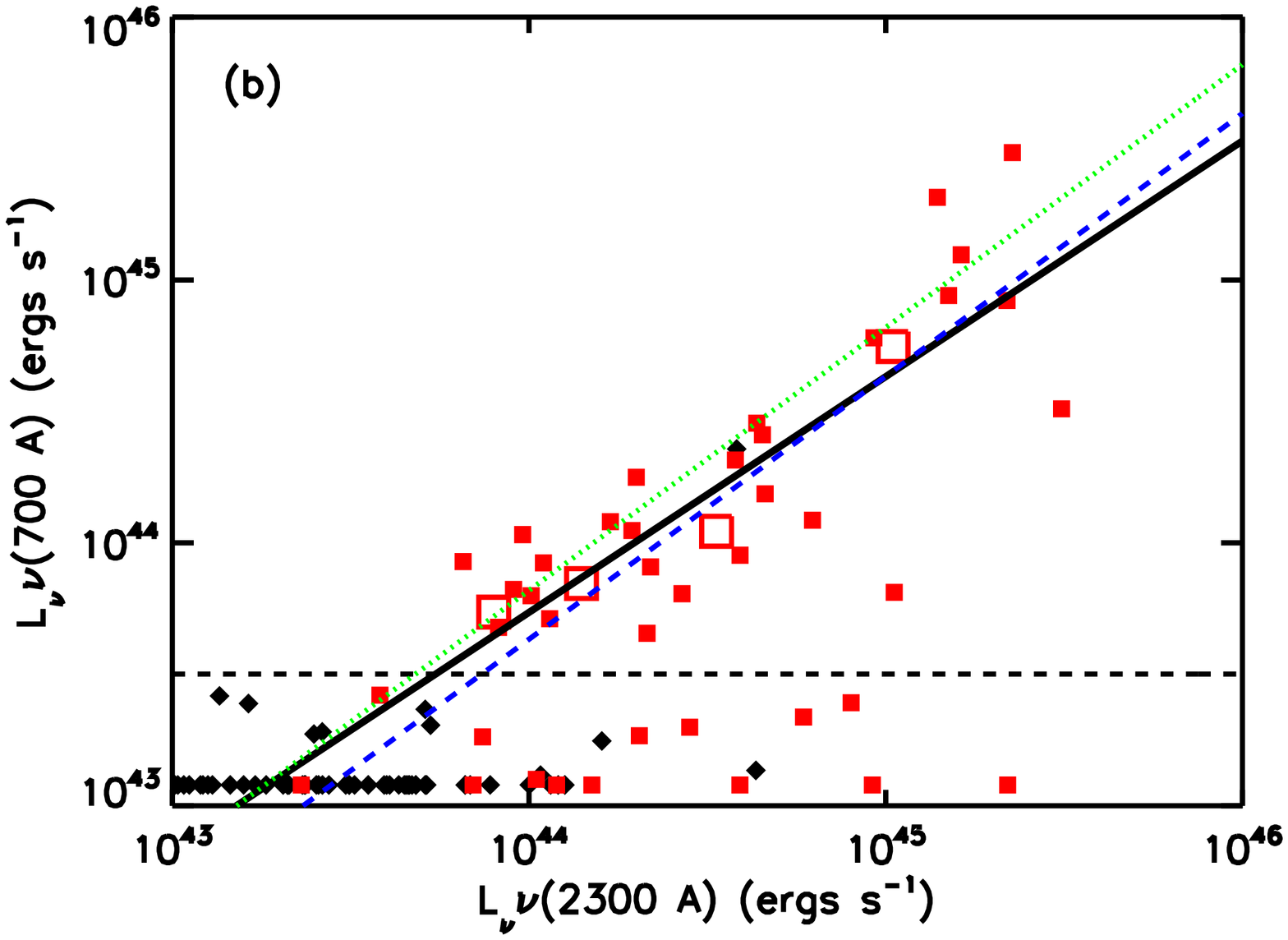,angle=90,width=3.8in}}
\centerline{\psfig{figure=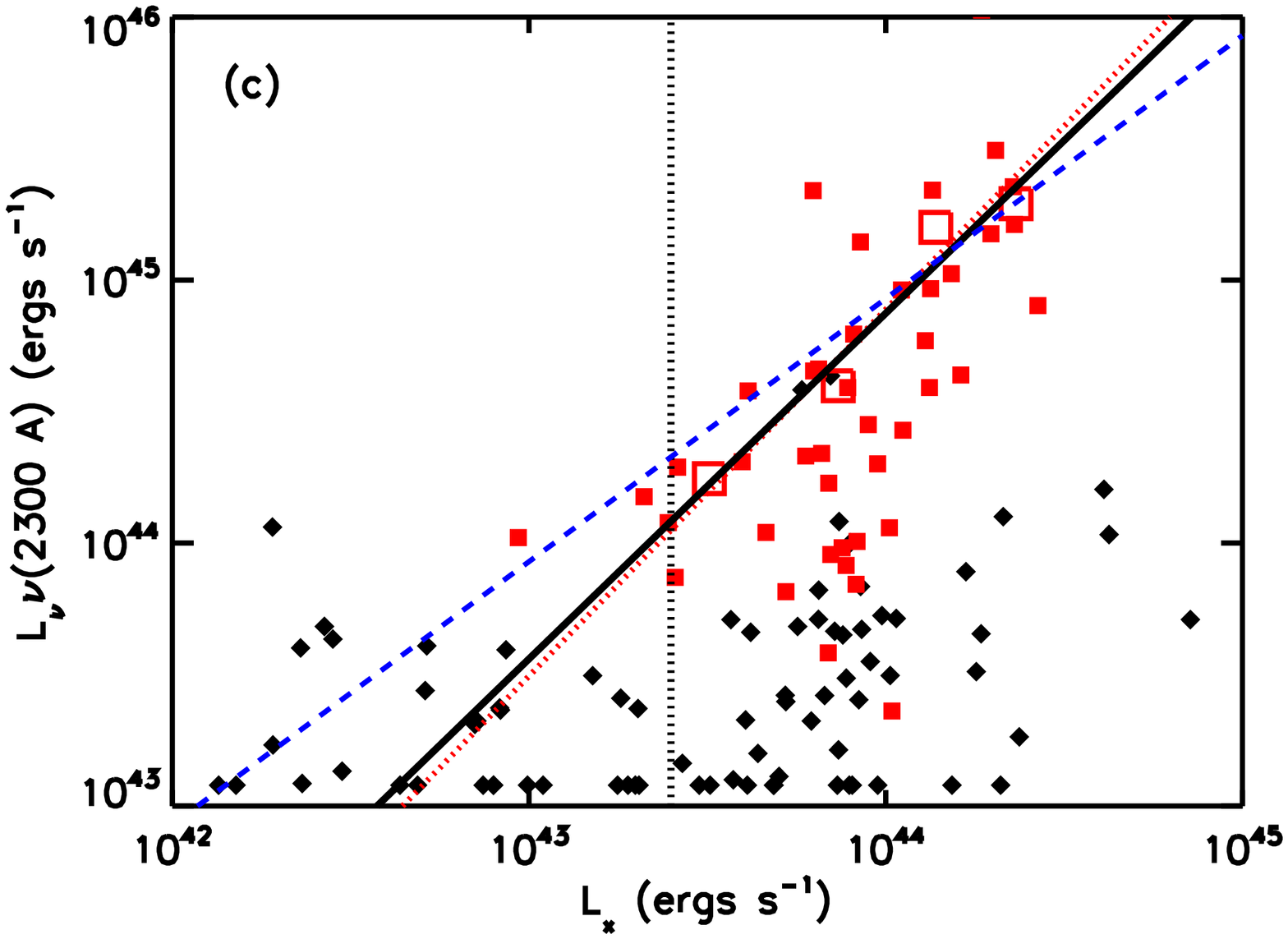,angle=90,width=3.8in}}
\figcaption[]{
The $z=0.9-1.4$ full X-ray sample
(broad-line AGNs - {\em red solid squares\/};
non--broad-line AGNs - {\em black diamonds\/};
means for the broad-line AGNs - {\em red open squares\/}).
(a) $L_\nu \nu$ at rest-frame 700~\AA, $L_\nu \nu$(700~\AA), 
vs. rest-frame $2-8$~keV luminosity, $L_X$. 
(b) $L_\nu \nu$(700~\AA) vs.  $L_\nu \nu$(2300~\AA). 
The green dotted line shows the relation for a source
with $L_{\nu}\sim\nu^{-1.35}$.
(c) $L_\nu \nu$(2300~\AA) vs. $L_X$. The red dotted line shows the 
relation converted from Vignali et al.\ (2003a).
In all three panels we show linear relations {\em (blue 
dashed lines)\/} and best power law fits {\em (diagonal black lines)\/}
(see text). In (a) and (b) the $3\sigma$ errors in $L_\nu \nu$(700~\AA) 
for the CLANS and CLASXS fields are shown by the dashed horizontal lines. 
In (a) and (c) the $L_X$ above which the field size 
corresponds to the wide area is shown by vertical dotted lines. 
Below this the sources are only in the much smaller area 
of the CDF-N. Sources with low or negative $y$-axis values are 
plotted at a nominal value of $1.2\times10^{43}$~ergs~s$^{-1}$. 
The only non--broad-line AGN with significant $y$-axis values in (a)
and (b) appears to be contaminated by a lower redshift galaxy (see text).
\label{lx_flux}
}
\end{figure*}

We can try to refine the results of \S\ref{seccont}
by considering how the ionizing
flux relates to other properties of the AGNs. In Figure~\ref{lx_flux}a
we show $L_\nu\nu$(700~\AA), where 700~\AA\ is the rest-frame 
wavelength, for the $z=0.9-1.4$ sources in the full X-ray 
sample versus rest-frame $2-8$~keV luminosity, $L_X$. We divide the 
sources into those with broad lines {\em (red squares)\/} and 
those without {\em (black diamonds)\/}. As might be expected, 
only the broad-line AGNs are ionizers. One of the non--broad-line 
AGNs is significantly detected, but an inspection of its spectrum 
shows that it appears to be contaminated by a lower redshift galaxy.
Thus, we exclude it from further consideration.

%
% FIGURE 9 (lx_lfuv and luv_lfuv and lx_luv)
%

In Figure~\ref{lx_flux}a we use large open squares to show the 
mean values of $L_\nu \nu$(700~\AA) versus $L_X$ for the 
$z=0.9-1.4$ broad-line AGNs. For all of the $z=0.9-1.4$ broad-line 
AGNs together we find a mean ratio of 
$L_{\nu}\nu{\rm (700~\AA)}/L_X=3.95$.
However, there is a very wide spread in the individual ratios. 
The dominant sources have ratios that are almost an 
order of magnitude higher than the mean ratio of 3.95.
The FUV and optical are more tightly correlated 
(see Fig.~\ref{lx_flux}b), so a considerable part of the observed
spread in Figure~\ref{lx_flux}a comes from the dispersion in the 
relation of the optical to the X-ray fluxes of the sources 
(see Fig.~\ref{lx_flux}c).
A linear relation with a normalization of 3.95 {\em (blue dashed line)\/} 
does not describe the mean values of $L_\nu\nu$(700~\AA) with $L_X$ 
very well. However, a non-linear relation does. 
The best power law fit to the mean values gives 
$L_{\nu}\nu{\rm (700~\AA)}/L_X=3.53\times 
(L_X/10^{44}~{\rm ergs~s}^{-1})^{0.46}$
with a $1\sigma$ uncertainty of 0.05 in the power law index
{\em (black solid line)\/}.
This non-linear dependence follows from the 
well-known non-linear dependences of the optical and ultraviolet
luminosities of broad-line AGNs on X-ray luminosity
(e.g., Vignali et al.\ 2003a, 2003b; Strateva et al.\ 2005;
Steffen et al.\ 2006; see Fig~\ref{lx_flux}c). 

In Figure~\ref{lx_flux}b we show the dependence of 
$L_\nu \nu$(700~\AA) on $L_\nu \nu$(2300~\AA) for the 
$z=0.9-1.4$ sources in the full X-ray sample, which
is tighter. The rest-frame $2300$~\AA\ luminosity is computed from
the $g$-band magnitudes obtained on the CFHT Megaprime camera
by Trouille et al.\ (2008). This filter is centered at
an observed-frame wavelength of 4900~\AA. Once again
we ignore small differential $K$-corrections assuming $f_\nu$
is flat. We see it is the broad-line 
AGNs that make up the UV-luminous population. The broad-line 
AGNs divide about equally into those which are well described
by a $L_{\nu}\sim\nu^{-1.35}$ power law between the two
wavelengths {\em (green dotted line)\/} and those which
have little or no ionizing flux. We use large open squares
to show the mean values of $L_\nu \nu$(700~\AA) versus 
$L_\nu \nu$(2300~\AA) for the $z=0.9-1.4$ broad-line AGNs. 
For all of the $z=0.9-1.4$ 
broad-line AGNs together we find a mean ratio of
$\nu L_{\nu}(700~{\rm \AA})/\nu L_{\nu}(2300~{\rm \AA})=0.47$.  
A linear relation with this normalization 
{\em (blue dashed line)\/} provides a reasonable fit to the 
mean values. In fact, the best power law fit to the mean 
values {\em (black line)\/}, which has a power law 
index of $0.90\pm0.18$, is consistent with the linear 
relation.

We can compare our mean ratio with that of Trammell et al.\ (2007),
who analyzed the GALEX fluxes for a large sample of Sloan
Digital Sky Survey (SDSS) quasars. Their quasars are about
an order of magnitude higher in luminosity
than the present sample, but their mean ratio of 
$\nu L_{\nu}$(700~\AA)$/\nu L_{\nu}$(2300~\AA) is similar. 
Trammell et al.\ (2007) only included sources with
detected FUV magnitudes, for which they find a ratio of about
0.5. Allowing for the roughly one-third of sources which
they do not detect, this value could be as low as 0.33. However,
given that these two numbers closely bracket our present results,
it is clear that there is not a large luminosity dependence
in the ratio. 

Finally, in Figure~\ref{lx_flux}c we show the well-known non-linear 
dependence of $L_\nu \nu$(2300~\AA) on $L_X$. The
best fit power law gives 
$\nu L_{\nu}(2300)/L_X = 7.46\times (L_X/10^{44}~{\rm ergs~s}^{-1})^{0.31}$
with a $1\sigma$ uncertainty of 0.21 in the index {\em (black line)\/}. 
This is almost identical to the relation found by 
Vignali et al.\ (2003a; see also Richards et al.\ 2005 for 
details on converting the Vignali et al.\ relation into the 
form used here) {\em (red dotted line)\/}.

One can next ask if there is any relation between the X-ray
properties of the broad-line AGNs and the ionizing photon escape.
In Figure~\ref{gam_type} we plot the ratio of 
$L_\nu \nu$(700~\AA)$/L_X$ for $L_X> 3\times10^{43}$~ergs~s$^{-1}$ 
sources with redshifts $z=0.9-1.4$ versus X-ray softness, as 
measured by the ratio of the 
$0.5-2$~keV flux to the $2-8$~keV flux. We divide the sources 
into those with broad lines {\em (red squares)\/} and those without 
{\em (black diamonds)\/}. We can again see
that all of the non--broad-line AGNs are UV faint. This immediately
tells us that we are not misidentifying sources
as non--broad-line AGNs when they are really broad-line AGNs,
as one might have expected to happen if the
broad lines were not visible spectroscopically due to dilution by
the host galaxy (Moran et al.\ 2002; Severgnini et al.\ 2003;
Cardamone et al.\ 2007). 
Barger et al.\ (2005) make a similar argument based on the 
presence of nuclei in the broad-line AGNs.

%
% FIGURE 10 (gam_type)
%
\begin{inlinefigure}
\psfig{figure=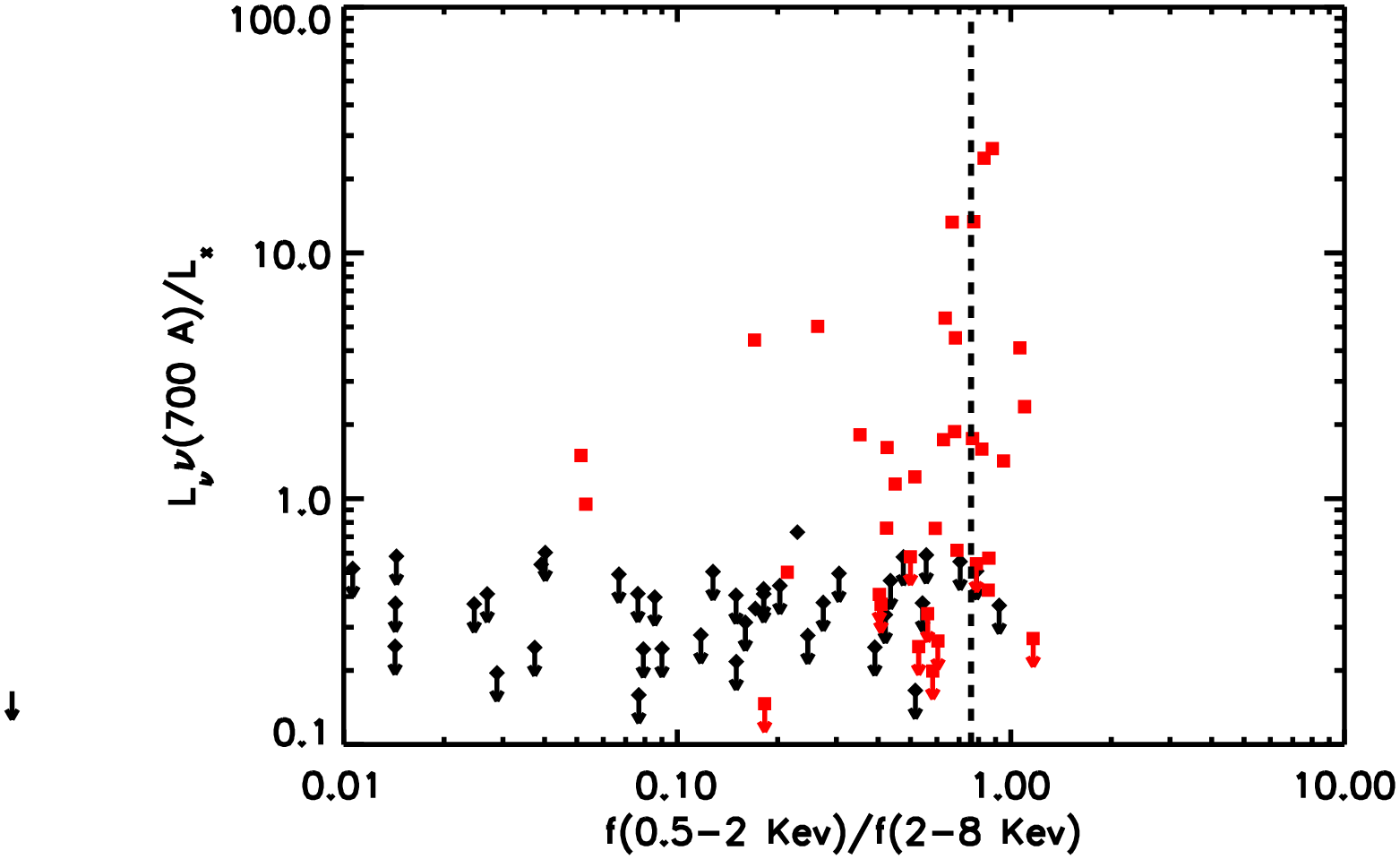,angle=90,width=3.8in}
\figurenum{10}
\figcaption[]{
Ratio of $L_\nu \nu$ at rest-frame 700~\AA, 
$L_\nu \nu$(700~\AA), to rest-frame $2-8$~keV luminosity, $L_X$, 
vs. X-ray softness, as measured by the ratio of the
$0.5-2$~keV flux to the $2-8$~keV flux, 
$f_{0.5-2~{\rm keV}}/f_{2-8~{\rm keV}}$ for the $z=0.9-1.4$
redshift range. Only sources with $2-8$~keV 
luminosities above $3\times 10^{43}$~ergs~s$^{-1}$ are shown. 
Broad-line AGNs are denoted by red squares, and non--broad-line 
AGNs by black diamonds. The flux ratio for a power law spectrum 
with photon index $\Gamma=1.8$ is shown by the vertical dashed line.
Sources with 700~\AA\ fluxes less than 0.2 $\mu$Jy are shown with
downward pointing arrows.
\label{gam_type}
}
\end{inlinefigure}

Indeed, we can pick out a large fraction of the broad-line AGNs
simply on the basis of their having a significant FUV flux.
Specifically, none of the 45 non--broad-line AGNs with
$L_X>10^{43}$~ergs~s$^{-1}$ 
and redshifts $z=0.9-1.4$
are detected above the FUV flux limit of $0.26~\mu$Jy ($3\sigma$),
while $70\pm14$\% of the broad-line AGNs are. Thus, the GALEX 
observations can be used to pick out most of 
the broad-line AGNs in the X-ray sample. More interestingly, the 
selection does not seem to depend on the X-ray softness, though the number
of X-ray hard broad-line AGNs in the sample is small. As can be seen
in Figure~\ref{gam_type}, we have six broad-line AGNs with
$0.5-2$~keV to $2-8$~keV flux ratios less than 0.3, and four of
these have significant FUV detections. There seems to be nothing
unusual about the optical spectra of the broad-line AGNs which are detected. 
Apparently, broad-line AGNs can have a high ratio of the ionizing flux to 
the X-ray flux, even if the source is quite
X-ray hard. This at once says that there is
not a one-to-one correspondence between the X-ray
spectral index and the neutral hydrogen opacity to the source, since
any substantial neutral hydrogen opacity would have the effect of
wiping out the ionizing flux, as indeed is invariably seen in
the non--broad-line AGNs. A detailed discussion of the comparison
between the X-ray and the optical and ultaviolet properties will
be given in L.~Trouille et al.\ (2008, in preparation). 

Regardless of the interpretation of the results,
we conclude that the production of the ionizing flux is 
best traced by optical spectroscopic broad-line AGN selection,
rather than by X-ray or combined X-ray and optical selection, 
since the X-ray color of the broad-line AGN
is not correlated with the presence
of the ionizing flux, and non--broad-line AGNs are simply not
detected. This considerably simplifies
the task of computing the evolution of the metagalactic
flux with redshift, since we believe that the spectroscopic 
broad-line AGNs are relatively completely identified in the 
X-ray samples.

The remaining question is why some broad-line AGNs are essentially
unobscured in the FUV (we shall refer to these as clear-channel 
broad-line AGNs), while others with similar luminosities
are substantially absorbed in the FUV (we shall refer to these as 
extinguished broad-line AGNs). A substantial part of the explanation is
obscuration by intervening LLSs along the line of sight which, even at $z=1$,
is a significant effect. At $z=1$ the density of LLS is approximately 0.9 per
unit redshift, which would result in about 40\% of the sources being obscured 
(e.g. Storrie-Lombardi et al.\ 1994). This is very similar to the fraction of 
sources with weak ionizing flux in the present sample and in 
Trammell et al.\ (2007).

However, a detailed analysis shows the situation is slightly 
more complicated. Storrie-Lombardi et al.\ (1994) find that 
Mg~II absorption line systems with rest-frame equivalent widths 
above 0.3~\AA\ are a very close proxy for LLS. Thus, we can 
approach the problem empirically, since we can easily detect 
such systems in the spectra of our broad-line AGNs. 

In Figures~\ref{uv_ratio}a and \ref{uv_ratio}b we show the 
spectral index $\alpha$ (where $L_\nu \sim \nu^{\alpha}$)
computed between 700~\AA\ and 2300~\AA\ versus
$L_\nu \nu$(2300~\AA). In Figure~\ref{uv_ratio}a we use
the observed FUV flux to compute the rest-frame 700~\AA\ flux, 
and in Figure~\ref{uv_ratio}b we use the observed NUV flux.
We do not include the additional $K$-correction as a
function of redshift relative to the mean redshift,  since
this would require assumptions about the SED. 
However, this effect should be small.
In Figure~\ref{uv_ratio}a we show the broad-line AGNs with $z=0.9-1.6$ 
and $L_\nu \nu$(2300~\AA$)>5\times10^{44}$~ergs~s$^{-1}$. 
In Figure~\ref{uv_ratio}b we show the broad-line AGNs with $z=1.9-2.5$ 
and $L_\nu \nu$(2300~\AA$)>2\times10^{45}$~ergs~s$^{-1}$. 
Note that the upper redshift limit in the latter is set by the 
requirement that Mg~II at the quasar redshift be on the 
spectrum. The spectral indices of sources with FUV fluxes 
below the $2\sigma$ threshold have been calculated using the 
$2\sigma$ threshold and are shown with downward pointing arrows.
At lower luminosities the uncertainties in the FUV fluxes
become too large to determine accurately the spectral indices. 
We use red squares to show sources without Mg~II absorbers
and we use black squares to show sources with Mg~II absorbers 
that are associated with the AGNs. We denote 
sources with intervening Mg~II absorbers along the
line of sight by black solid diamonds when the
redshift of the system is such that it would substantially
quench the FUV light and by black open diamonds
when the redshift of the system is such that it only 
partially covers the wavelengths observed by the FUV filter.

As expected, the presence of an intervening Mg~II absorber at a redshift
which would substantially quench the FUV light (the
black diamonds) guarantees 
that a source is extinguished. (We take sources with spectral 
indices less than $-2$ to be extinguished.) 
However, two of the clearly extinguished
sources in the lower redshift interval 
(Fig.~\ref{uv_ratio}a) do not have Mg~II absorbers. 
In one case the wavelength coverage of the spectrum does not
extend to short enough wavelengths to ensure that there
is not a LLS along the line of sight that could be significantly
extinguishing the flux in the FUV filter. In the other
case there is no Mg~II absorption, and if the extinction is 
caused by a LLS, then it must have weak Mg~II absorption. The 
possible very small fraction of LLS without Mg~II absorption 
is discussed in Storrie-Lombardie et al.\ (1994) and 
Steidel \& Sargent (1992).

%
% FIGURE 11 (uv_ratio and uv_ratio2)
%
\begin{inlinefigure}
\psfig{figure=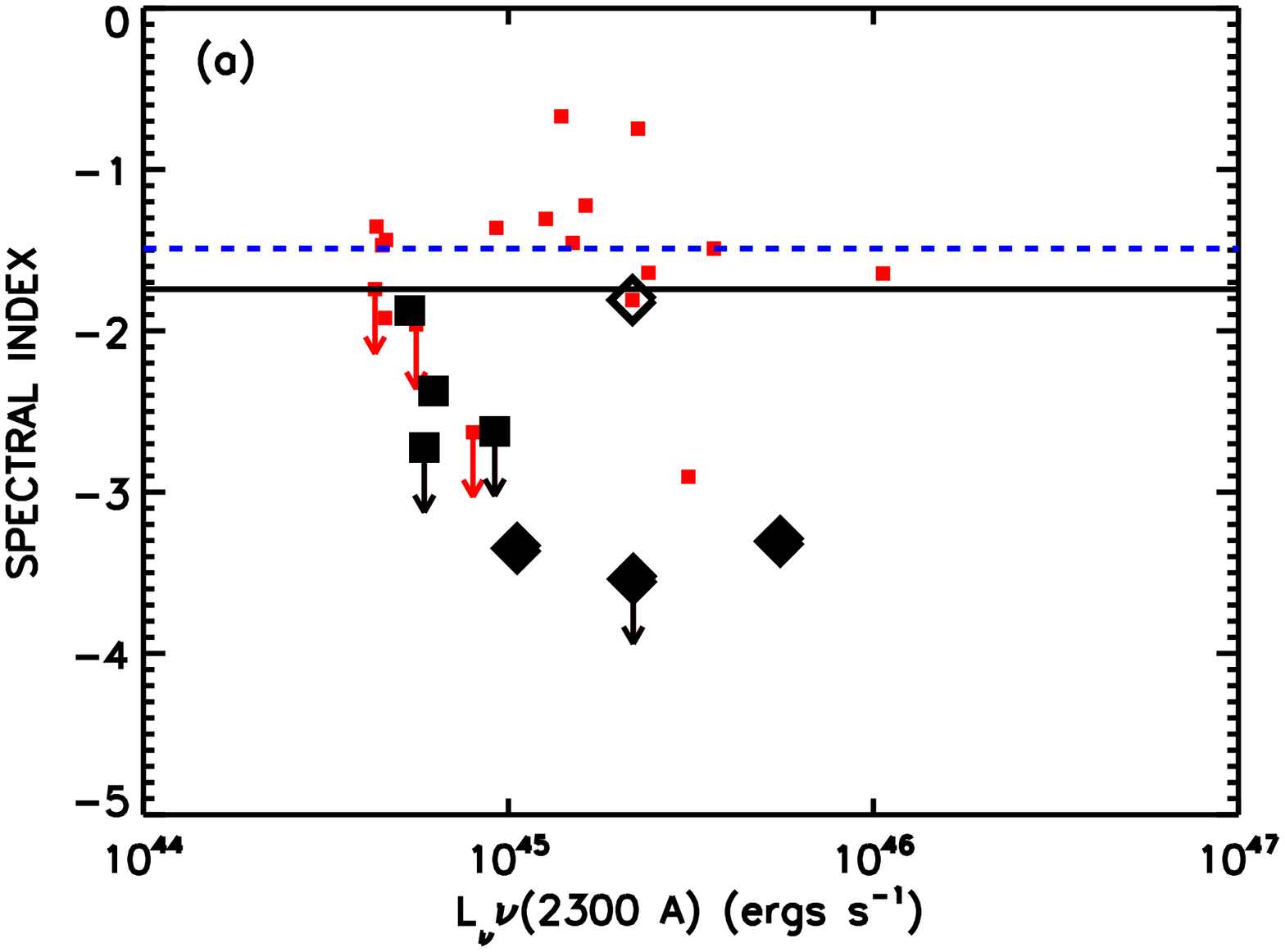,angle=90,width=4.0in}
\psfig{figure=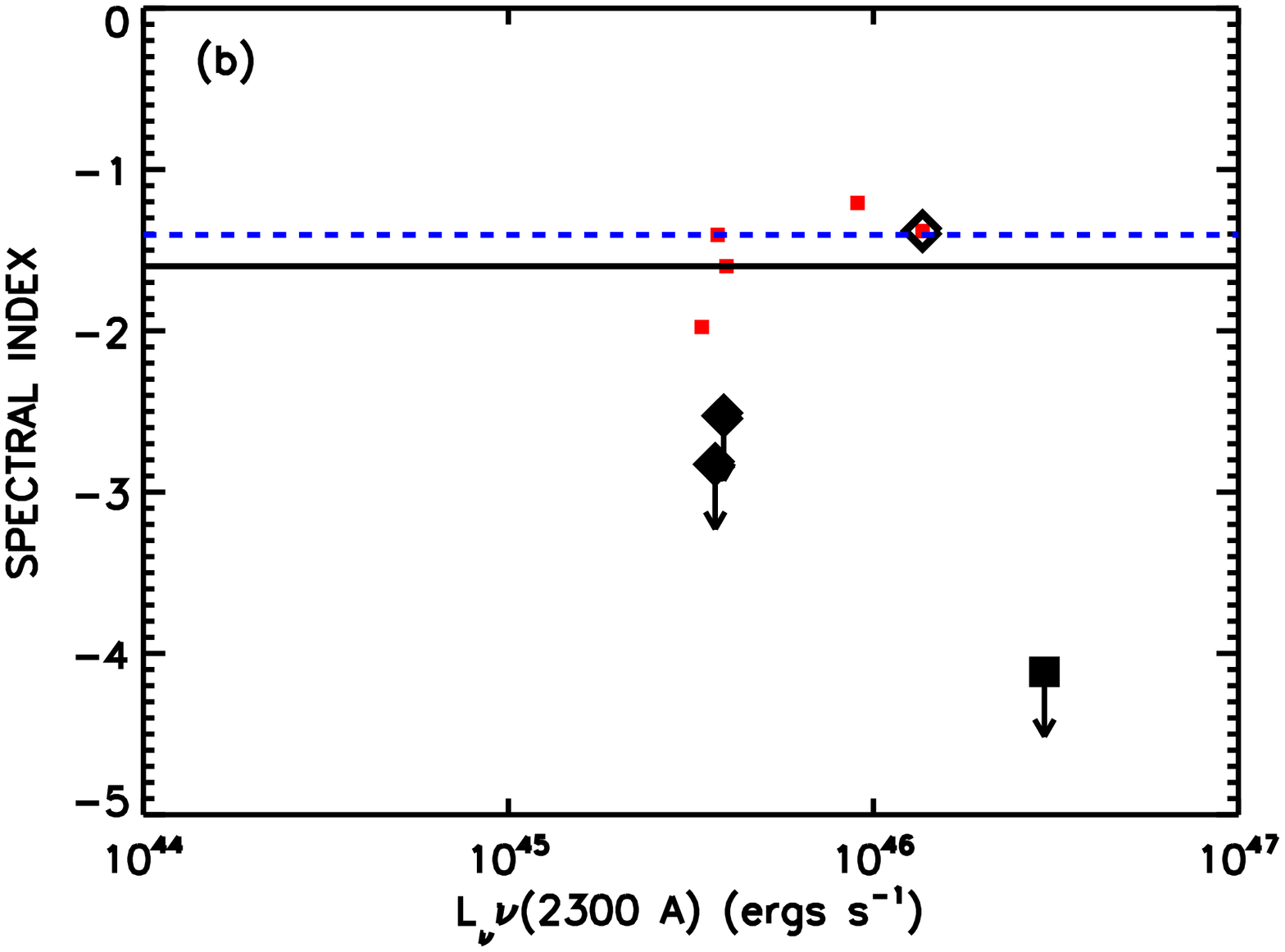,angle=90,width=4.0in}
\figurenum{11}
\figcaption[]{
(a) Spectral index computed between 700~\AA\ and 2300~\AA\
for $z=0.9-1.6$ broad-line AGNs with rest-frame 2300~\AA\ 
luminosities above $5\times10^{44}$~ergs~s$^{-1}$ vs. $L_\nu \nu$ 
at rest-frame 2300~\AA, $L_\nu \nu$(2300~\AA).
Sources with 700~\AA\ fluxes below the $2\sigma$ threshold are shown 
at the $2\sigma$ threshold with downward pointing arrows. 
Sources without Mg~II absorbers in the line of sight are shown 
with red small squares. Sources 
with Mg~II absorbers associated with the AGNs are shown with 
black larger squares. Sources with intervening Mg~II absorbers 
along the line of sight are shown with black solid diamonds when 
the redshift of the system is such that it would substantially 
quench the FUV light and with black open diamonds when the
redshift of the system is such that it would only partially 
cover the wavelengths observed by the FUV filter. The black 
line shows the median spectral index 
for all of the sources. The blue dashed line shows the median
spectral index for the sources after removing those
with spectral indices less than $-2$ that do not have Mg~II
absorbers associated with the AGNs.
(b) Same plot for $z=1.9-2.5$ broad-line AGNs with rest-frame
2300~\AA\ luminosities above $2\times 10^{45}$~ergs~s$^{-1}$.
\label{uv_ratio}
}
\end{inlinefigure}

Interestingly, for four of the sources with Mg~II absorbers
in Figure~\ref{uv_ratio}a, the Mg~II 
system is a narrow-line system associated with the AGN 
(within 10,000~km~s$^{-1}$ of the AGN) and is not intervening
{\em (the four black squares)\/}. 
In three of the cases the velocity separation is less than 
2000~km~s$^{-1}$ from the associated AGN. For one other source 
in Figure~\ref{uv_ratio}b {\em (the black square)\/}
the quasar shows broad absorption features. 
All of these cases show some level of absorption
but in two cases we still see an ionizing flux.
Either the conversion of MgII to neutral hydrogen in
these systems is different from that of the intergalactic
sytems or they only partially cover the ionizing region.
The actual explanation is not important for the present purposes 
since there is a critical 
distinction between associated and intervening systems.
In the associated cases the ionizing radiation is
absorbed before it emerges into the general intergalactic medium.
In contrast, intervening systems are
part of the radiative transfer in the intergalactic medium.
Thus, in computing the ionizing source term we should not correct for
absorption by the associated systems while we should remove the effects
of absorption in the intervening systems. 

The median spectral index of all of the sources in the 
lower redshift interval is shown by the black line 
in Figure~\ref{uv_ratio}a and has a value of $-1.8$ $(-2.7, -1.4)$,
where the bracketed indices give the 68\% confidence range
computed using the median sign method. If we remove
the three systems with substantial intervening absorption
{\em (three black solid diamonds)\/}
and the two clearly extinguished sources without Mg~II absorption
{\em (two red squares)\/},
assuming they are also caused by an intervening 
LLS, then the median spectral 
index becomes $-1.45$ $(-1.9, -1.2)$ {\em (blue dashed line)\/}
and the mean spectral index is $-1.35$.
The number of sources in the higher redshift interval 
(Fig.~\ref{uv_ratio}b) is small, but the median spectral index 
after removing the two systems with substantial intervening 
absorption {\em (two black solid diamonds)\/}
is consistent with that in the lower redshift interval. 
Thus, we assume the mean intrinsic spectral 
index is redshift invariant and adopt
the mean value of $-1.35$ at $z=1.15$ in computing
the evolution of the metagalactic ionizing flux in \S\ref{secmet}.
Our 700~\AA\ to 2300~\AA\
flux ratio is 0.20, and we scale this to a 912~\AA\ to 2300~\AA\
flux ratio of 0.26, assuming an intrinsic spectrum with an index
of $-1$ below the break, which is similar to that of the 
broad-line AGNs above the break. This is a factor of 1.5 lower 
than the value adopted in Haardt \& Madau (1996). Using our
value in their analysis would bring their results closer
to the present values.

\section{Evolution of the Metagalactic Ionizing Flux}
\label{secmet}

The discussion of the previous sections suggests that
the best procedure to measure the metagalactic
flux is to take the broad-line AGN sample
from the X-ray data and to measure its light at NUV wavelengths
above the Lyman break. We can then use the average spectral
break, corrected for intergalactic absorption, to derive
the ionizing flux emissivity. Our broad-line AGN sample
chosen from the full X-ray sample probes to much fainter UV
magnitudes than the large optical sample and therefore
avoids major extrapolations in determining the UV volume
emissivity at high redshifts. Using the measured UV fluxes
rather than normalizing to the X-ray fluxes avoids
the scatter and non-linearity in the relation of the
UV to the X-ray. Finally, measuring the flux at the
longer wavelengths and then correcting to the ionizing
flux avoids the complex problem of correcting for the
intervening intergalactic absorption, though at the expense
of assuming a redshift invariant average spectrum for the
broad-line AGNs.

In Figure~\ref{ion_evol} we show the evolution of the comoving
volume emissivity determined in this way from the
broad-line AGNs in the CLANS, CLASXS and CDF-N fields
{\em (red squares)\/}. We first 
computed the rest-frame 2300~\AA\ emissivity for our broad-line AGNs 
in each redshift interval. To determine the rest -frame 2300~\AA\ flux, we 
interpolated (or extrapolated at the the highest redshifts) using
the NUV magnitudes and the 
$g$ (observed frame 4900~\AA) and $i$ (observed-frame 7900~\AA) 
magnitudes from Trouille et al.\ (2008).
We then scaled the 2300~\AA\ emissivity to the ionizing volume emissivity
using the results from \S\ref{secxrayion}. The errors were again
computed with the jackknife method. In order to test the completeness,
we computed the results as a function of the limiting X-ray flux.
Below $z=2.5$ the results are dominated by sources with X-ray
fluxes above $7\times10^{-15}$~ergs~cm$^{-2}$~s$^{-1}$, which are
sampled by the larger area. Above this redshift the results
are dominated by sources in the smaller CDF-N area and cosmic
variance could be a more significant issue. However, the results
substantially converge at fluxes well above the limiting flux
in the CDF-N sample. We do not detect any broad-line AGNs
above $z=4$, implying a large fall-off in the ionizing fluxes above $z=4$.
The evolution of our comoving volume emissivity can 
be parameterized by the equation
\begin{equation}
\lambda_{ion}=(1.67+15.91 z-3.660 z^{2}) 10^{23}\ 
{\rm ergs}\ {\rm s}^{-1}\ {\rm Hz}^{-1} {\rm Mpc}^{-3}\,,
\label{eqnlam}
\end{equation}
which is shown as the solid red curve in Figure~\ref{ion_evol}.

%
% FIGURE 12 (ion_evol)
%
\begin{inlinefigure}
\centerline{\psfig{figure=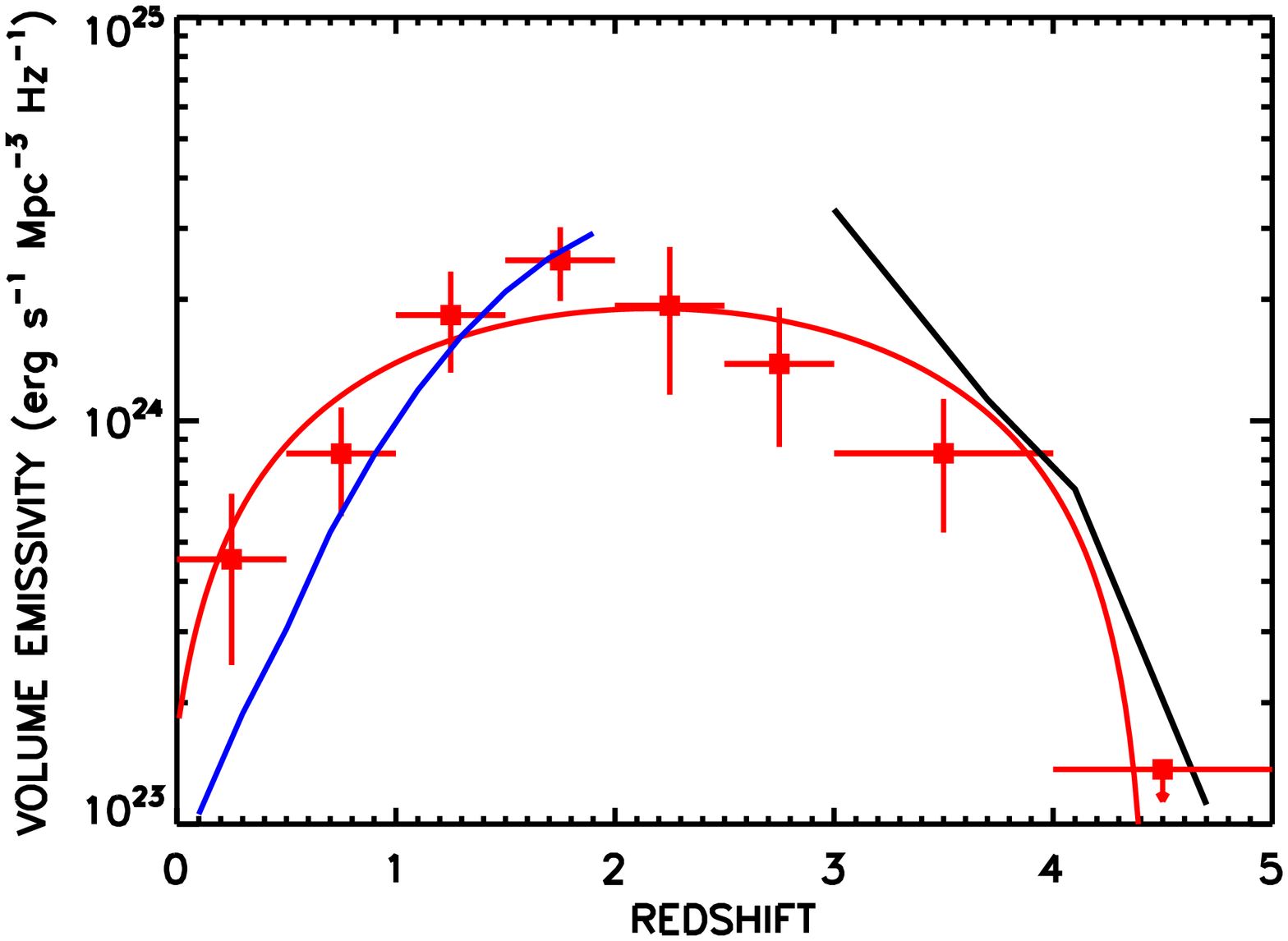,angle=90,width=4.0in}}
\figurenum{12}
\figcaption[]{
Redshift evolution of the comoving volume
emissivity of the ionizing flux just below the Lyman continuum 
edge from broad-line AGNs. 
The red squares show the values calculated directly from 
the broad-line AGNs in the full X-ray sample (see text). 
The error bars are $1\sigma$ 
calculated using the jackknife method. There are no broad-line 
AGNs above $z=4$ in our sample, and we show the upper limit 
based on assuming a one sigma upper limit of 1.8 sources
(Gehrels 1986) with the characteristic 
break luminosity shown in Figure~\ref{lum_evol}b.
The blue curve was calculated from the maximum 
likelihood fit to the Richards et al.\ (2005) optically selected 
broad-line AGN sample. We show this only over its valid redshift 
range. The black solid curve shows the source function computed 
by Meiksin (2005), which was estimated from the SDSS survey data 
for broad-line AGNs combined with lower luminosity broad-line 
AGN constraints from an optical survey at $z=3$. We have corrected 
this to the presently derived spectral shapes. 
\label{ion_evol}
}
\end{inlinefigure}

In the above calculation we have worked directly with the 2300~\AA\ emissivities 
determined from the broad-line AGN sample in the  CLANS, CLASXS, and CDF-N 
fields. A similar result can be derived from broad-line AGN X-ray luminosity functions 
determined as a function of redshift (e.g., Yencho et al.\ 2008 and references therein), 
but this would require the use of the more uncertain conversion from X-ray emissivity
to ionizing emissivity.

We can also determine the results from the luminosity functions
of optically selected samples of broad-line AGNs, where these are 
deep enough. Here we have computed the comoving ionizing volume 
emissivity of the broad-line AGNs from the maximum
likelihood fit to the Richards et al.\ (2005) 
optically selected sample {\em (blue curve)\/}. 
As would be expected from the
fact that Richards et al.\ (2005) found their luminosity 
functions to be in almost perfect agreement with the X-ray 
luminosity functions, the results derived from the optical
and X-ray samples are in extremely good agreement over 
$z=0-2$, where the optical samples are sufficiently deep 
to make this comparison. It is much harder to use the 
existing wide-field optically selected samples of 
broad-line AGNs to make such a comparison at high redshifts, 
since such samples are not deep enough to probe the dominant 
contributors.  Extrapolations outside the range for which 
a fit has been derived can produce wildly inaccurate results
as discussed by for example Richards et al. (2006). 

Meiksin (2005) used the SDSS observations of broad-line AGNs
together with the lower luminosity broad-line AGN constraints from 
the optical survey of Steidel et al.\ (2002) and Hunt et al.\ (2004)
at $z=3$ to estimate the 
higher redshift values from the optical samples. 
We show his results with the black solid curve. (We have
modified the Meiksin 2005 result which assumed a slightly
different value of the break across the Lyman edge to the presently derived
spectral shape.) 
Given the extrapolation uncertainties
in this type of estimate, the agreement is quite good.
As is well known, the results of Haardt \& Madau (1996)
and Madau et al.\ (1999) are too high because
of the assumed quasar luminosity functions. However,
the present results are considerably lower than even
the modified Haardt-Madau values that are often used
(e.g., Haardt \& Madau 2001; Bolton et al.\ 2005).

As we have discussed, the production of the ionizing flux 
is dominated by a small number of sources. We can investigate this
directly using the present X-ray selected sample. Typically
$3-10$ sources produce most of the flux in our line of sight in the redshift 
intervals of Figure~\ref{ion_evol}, even with a 1~deg$^2$ 
area. (Different lines of sight would see different
AGNs as being broad-line and having significant ionizing flux
because of the geometric effects of the intrinsic absorption
in the AGNs. However, the bulk of the luminous X-ray sources
are broad-line AGNs [e.g., Steffen et al. 2003; Barger et al. 2005],
and so we expect this effect to be small.) We quantify the number of
contributors in Figure~\ref{lum_evol}a, where we 
show the number density of sources producing 67\% of the 
observed ionizing flux. The number density drops from a value of 
$10^{-5}$~Mpc$^{-3}$ at low redshifts, where lower luminosity
sources dominate, to about $10^{-6}$~Mpc$^{-3}$ at
$z=3.5$. In Figure~\ref{lum_evol}b we show the
luminosity range that produces from $25-75$\% of the
ionizing flux. This emphasizes that a rather
narrow range of luminosities straddling the break
luminosity in the luminosity function produces most of the flux. The typical
luminosity of the dominant sources rises quite rapidly
at low redshifts, reflecting the rapid nearly pure luminosity
evolution of the optical and X-ray luminosity functions
in this redshift range 
(e.g., Barger et al.\ 2005; Richards et al.\ 2005).
It then reaches a fairly invariant characteristic
luminosity of about $7\times10^{29}$~ergs~s$^{-1}$~Hz$^{-1}$
above this redshift. This corresponds to an absolute
rest-frame magnitude at 2300~\AA\ of M$_{2300}=-24.5$.

%
% FIGURE 13 (num_evol and lum_evol)
%
\begin{inlinefigure}
\vskip 0.5cm
\centerline{\psfig{figure=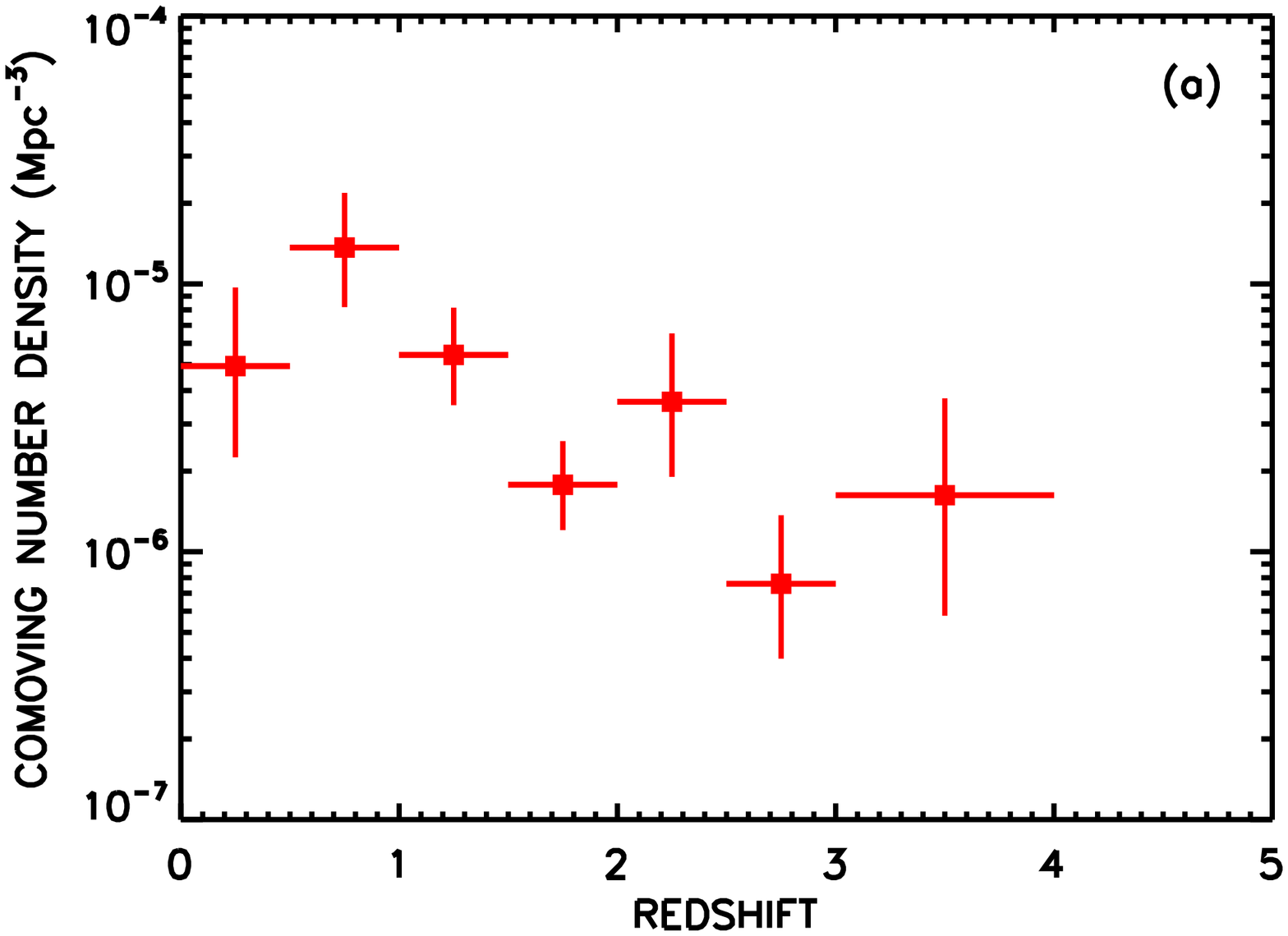,angle=90,width=3.3in}}
\centerline{\psfig{figure=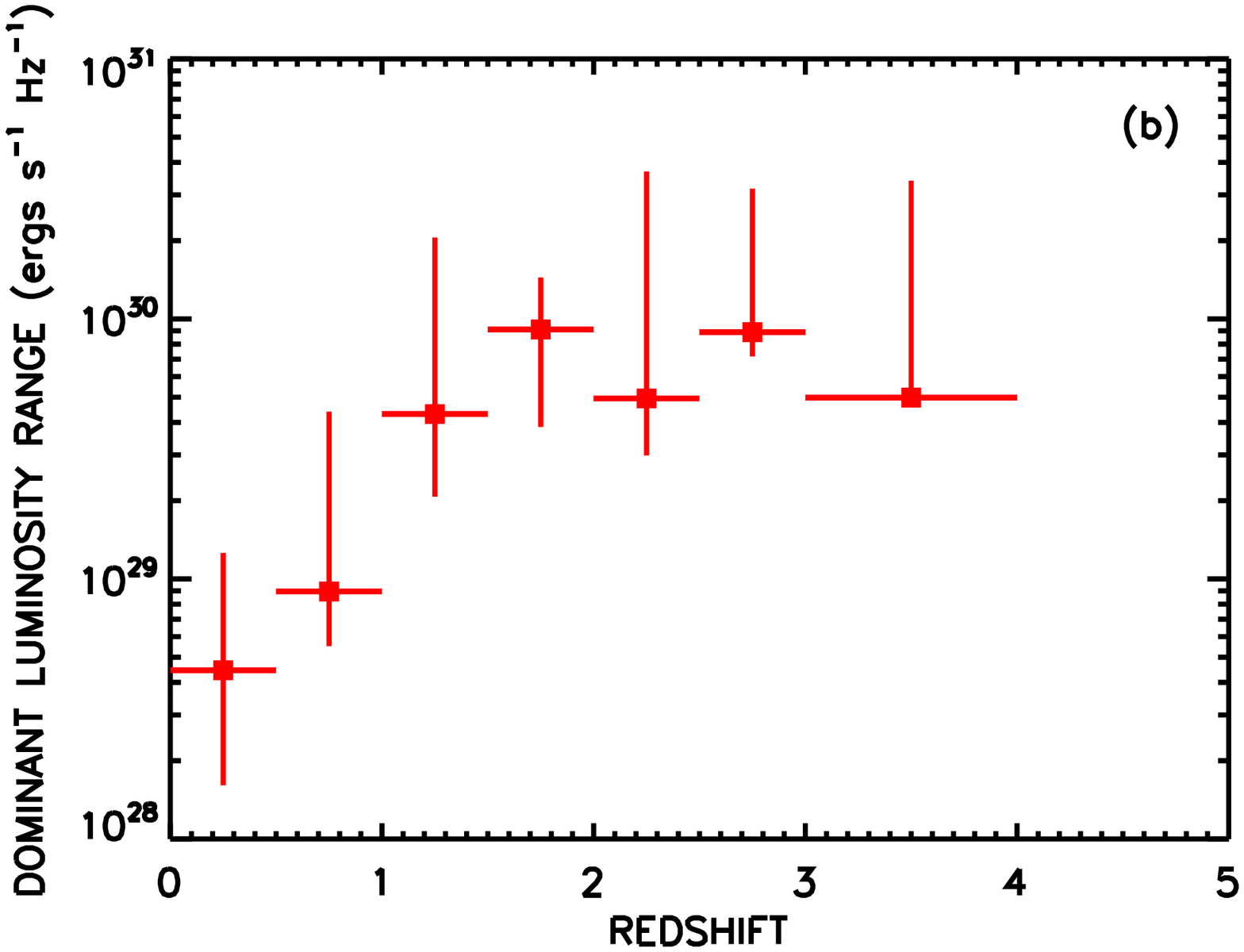,angle=90,width=3.3in}}
\figurenum{13}
\figcaption[]{
(a) Redshift evolution of the comoving volume
density of the sources producing 67\% of the ionizing volume
emissivity. The $1\sigma$ error bars have been computed
using the median sign method. The number of dominant sources
in each redshift interval ranges from 3 to 10.
(b) The luminosity range of the sources dominating the
ionizing volume emissivity. The squares show the luminosity
above which 50\% of the emissivity is produced, and the
vertical lines show the 25\% to 75\% range.
\label{lum_evol}
}
\end{inlinefigure}

As pointed out by Madau et al.\ (1999), the measured absorption
distance for the ionizing photons at $z\sim3$ is quite
small so that the ionization rate is only determined
by local sources. They estimate an absorption distance
corresponding to $\Delta z= 2.8 (1+z)^{-2}$ at these
redshifts, which is $\Delta z=0.18$ at $z=3$.
Combining this with the comoving number density in
Figure~\ref{lum_evol} gives a contributing number of
quasars of about 3000 at $z=3$ and 200 at $z=4$ in
any given region. Beyond
this the numbers drop rapidly, and at $z=4.5$ there would
be less than 20 dominant quasars. We may compare this with
Meiksin \& White (2003), who self-consistently estimated
the absorption distance by normalizing the modeled
quasars to match the required ionization in the Ly$\alpha$ 
forest and then computing the absorption distance, rather
than obtaining it from observations.
The number of contributing quasars is higher by factors of 
several, as would be expected since the present AGN emissivity 
is too low to account for the ionization at these redshifts. 
As the number of contributing quasars drops,
fluctuations become important, and the contribution of the
AGNs to producing transmission in the AGNs is further reduced,
increasing the deficiency of these sources for ionizing the
IGM (Meiksin \& White 2003).

%
% FIGURE 14 (ion_evol_opt)
%
\begin{inlinefigure}
\centerline{\psfig{figure=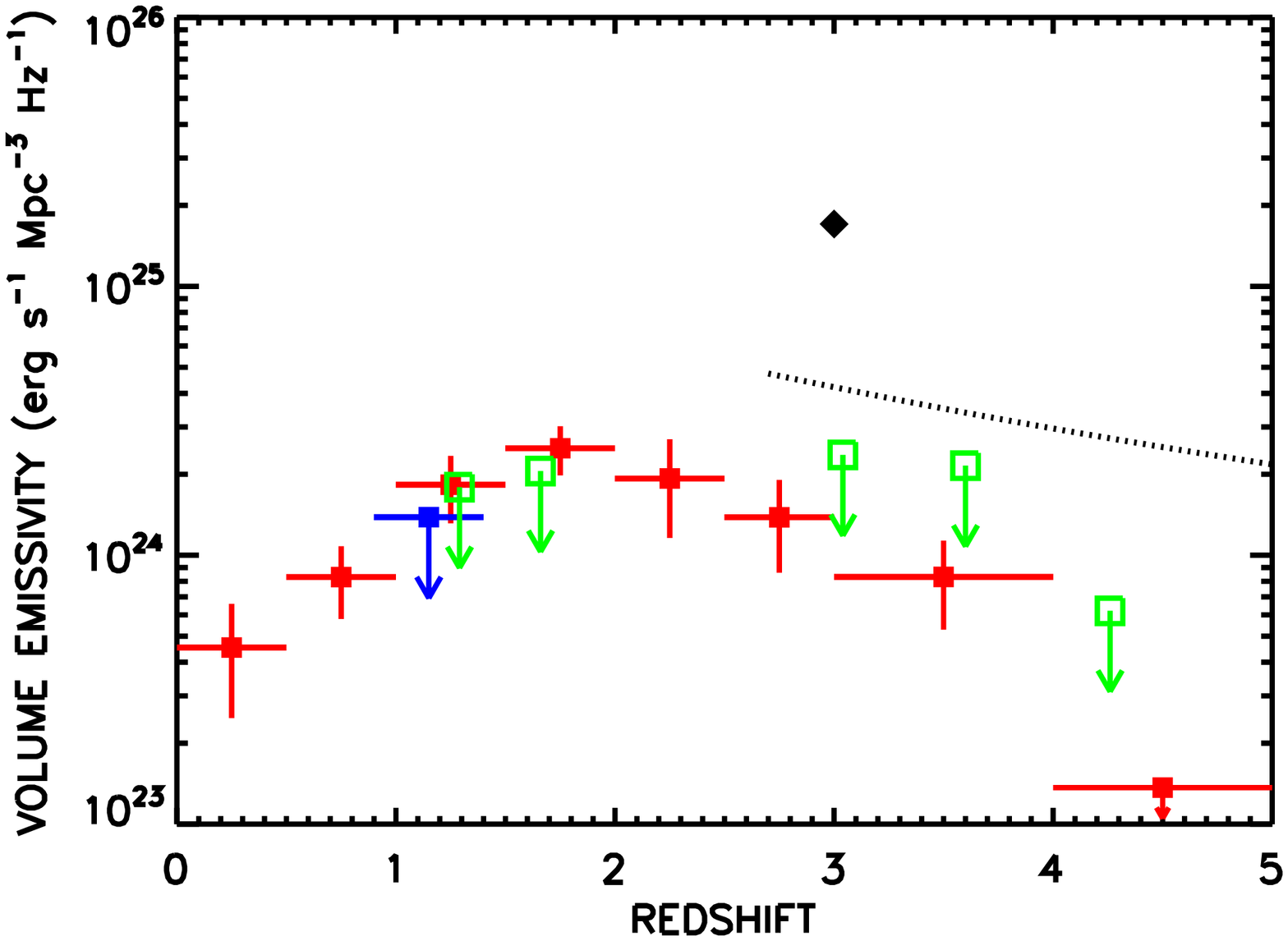,angle=90,width=4.0in}}
\figurenum{14}
\figcaption[]{
Redshift evolution of the comoving volume emissivity of the 
ionizing flux just below the Lyman continuum break calculated from
the broad-line AGNs in the full X-ray sample
{\em (red squares)\/} compared with the upper limits 
for galaxies {\em (blue and green squares)\/}.
The blue solid square shows our directly measured $2\sigma$
limit at $z=1.15$. The green open squares show the $2\sigma$
limits obtained by converting the 1500~\AA\ comoving emissivities
from Tresse et al.\ (2007) to the ionizing emissivities
using our $2\sigma$ upper limit on the average ionization
fraction at $z=1.15$ (see \S\ref{secoptsam})
and then multiplying by 1.26 to correct from 700~\AA\ to 912~\AA.
The black diamond shows the measured value from Shapley et
al.\ (2006)'s LBGs. The dotted line shows the required emissivity
to reproduce the Ly$\alpha$ forest structure in the 
IGM estimated by Meiksin (2005).
\label{ion_evol_opt}
}
\end{inlinefigure}

In Figure~\ref{ion_evol_opt} we compare the comoving volume
emissivity of the ionizing light just below the Lyman continuum edge 
calculated from the broad-line AGNs in the full X-ray sample {\em (red squares)\/} 
with the upper limits for galaxies {\em (blue and green squares)\/}.
At $z=1.15$ we show the directly
measured $2\sigma$ upper limit from the present work
as the blue square with the downward pointing arrow.
To compute the limits at higher redshifts,
we converted the 1500~\AA\ comoving emissivities from
Tresse et al.\ (2007) to the ionizing 
emissivities using our $2\sigma$ upper 
limit on the average ionization fraction at $z=1.15$ 
(see \S\ref{secoptsam}) and then
multiplied by 1.26 to correct from 700~\AA\ to 912~\AA. 
For the highest redshift points ($z \gtrsim 3$) we used the unweighted
estimated values from Tresse et al.\ (2007)'s Table~4.
We show these limits on the comoving ionizing emissivity derived from
the Tresse et al.\ (2007) data as green open squares
with downward pointing arrows in Figure~\ref{ion_evol_opt}.
The Tresse et al.\ (2007) 1500~\AA\
emissivities are broadly consistent with previous 
measurements (Steidel et al.\ 1999; Gabasch et al.\ 2004; 
Arnouts et al.\ 2005; Sawicki \& Thompson 2006) and
with the present work.
The black dotted curve in Figure~\ref{ion_evol_opt} shows the
the required emissivity to reproduce the Ly$\alpha$ forest structure in the
intergalactic medium at $z>3$ estimated by Meiksin (2005).
We have used the value allowing for reprocessing of the
ionizing radiation in the IGM (the dashed curve of Meiksin's
Figure~1).

If the high-redshift ionization fraction is as low as our
measured limit at $z=1.15$, then the upper limit on the galaxy contribution 
is slightly low compared with the required ionization.
This result is marginal in the $z=3-4$ range but stronger
above $z=4$. However,
given the rapid change in the morphologies of the galaxy population
beyond $z=1$, we may question whether the $z=1.15$ ionization fraction 
of the present work is applicable to the higher redshift population.
If the ionization fraction rises to that measured 
by Shapley et al.\ (2006) from their small sample of LBGs, 
then the galaxy contribution at high redshifts, as shown
by the black diamond at $z=3$, would easily match (or indeed
exceed) the required ionization. 
(We  note that if Iwata et al.\ [2008] are correct
in their null detection of the brighter of the two Shapley et al.\ [2006]
LBG detections, then the Shapley et al.\ point would fall by a factor of 
roughly 3.) However, the Shapley et al.\ (2006) sample may be 
biased towards the LBGs that are most likely to have ionizing 
flux and thus may really be an upper limit even on the LBG contribution. 
It is therefore critical to understand 
whether the current measurements of the $z\sim 3$ LBGs are 
representative of the population as a whole at these redshifts. 
The narrow-band ionizing flux surveys of $z\sim3$ LBGs currently
being worked on by several groups should resolve this question.

\section{Summary}
\label{secsummary}

We have used X-ray, optical, and GALEX observations to measure the 
contribution of AGNs to the ionizing flux as a function of redshift. Our 
analysis of a large population of X-ray sources confirms that
the AGN contribution to the ionizing flux peaks at around 
$z=2$ and drops rapidly at higher redshifts. It is insufficient 
to account for the observationally inferred ionizing flux at 
high redshifts ($z>3$). We have also obtained a strong upper 
limit on the contribution of galaxies to the ionizing flux at 
$z=1.15$ using GALEX observations
of a very large sample of rest-frame UV selected galaxies
in the GOODS-N region. Our $2\sigma$ upper limit on the 
ionization fraction for this population, 
$f_\nu(700~{\rm \AA})/f_\nu(1500~{\rm \AA})=0.008$,
yields an upper limit on the comoving ionizing emissivity
from the galaxies at $z=1.15$, which 
is at most comparable to that for the AGNs at $z=1.15$.
We briefly discuss the possibility that the galaxies may
in fact be net absorbers at these redshifts.
 
If galaxies are to contribute significantly to the ionizing radiation, 
then the ionization fraction must increase at higher redshifts.
Shapley et al.\ (2006)'s measured value from their small
sample of LBGs at $z=3$ is about six times higher than the value
from our $2\sigma$
upper limit on the ionization faction at $z=1.15$. However,
the Shapley et al.\ (2006) value would be reduced by about 
a factor of three if the object which is not confirmed
by Iwata et al.\ (2008) is removed. If these LBGs are representative of the 
high-redshift population as a whole, then this would be adequate 
to produce the high-redshift ionizing flux.
However, it would require an increase in the 
escape of the ionizing photons from the galaxies
as a function of increasing redshift.

\acknowledgments
We thank the anonymous referee for comments that substantially improved the
manuscript. We thank Dan Vandenberk and Avery Meiksin for helpful
discussions. We gratefully acknowledge support from NSF grants
AST-0407374 and AST-0709356 (L.~L.~C.) and
AST-0239425 and AST-0708793 (A.~J.~B.),
the University of Wisconsin Research Committee with funds 
granted by the Wisconsin Alumni Research Foundation, 
and the David and Lucile Packard Foundation (A.~J.~B.).
L.~T. was supported by an NSF Graduate Fellowship.

\end{document}